\documentclass[onecolumn,showpacs,amsmath,amssymb,11pt]{revtex4}
\input{epsf}

\usepackage{CJK}
\usepackage{graphicx}
\usepackage{array}
\usepackage{dcolumn,longtable}
\usepackage{lscape}
\usepackage{rotating,booktabs}
\usepackage{booktabs,threeparttable}
\begin{document}

\title{\bf General theory for Rydberg states of atoms: nonrelativistic case}
\author{Xiao-Feng Wang$^{1,2}$, Zong-Chao Yan$^{2,3,4}$}

\affiliation{$^1$ School of Physics and Technology, Wuhan University, Wuhan 430072, P. R. China}

\affiliation{$^2$ Department of Physics, University of New
Brunswick, Fredericton, New Brunswick, Canada E3B 5A3}

\affiliation {$^3$ State Key Laboratory of Magnetic Resonance and
Atomic and Molecular Physics, Wuhan Institute of Physics and
Mathematics, Chinese Academy of Sciences, Wuhan 430071, P. R. China}

\affiliation {$^4$ Center for Cold Atom Physics, Chinese Academy of
Sciences, Wuhan 430071, P. R. China}

\date{\today}

\begin{abstract}
We carry out a complete derivation on nonrelativistic energies of atomic Rydberg states, including finite nuclear mass corrections.
Several missing terms are found and a discrepancy is confirmed in the works of
Drachman [in \emph{Long Range Casimir Forces: Theory and Recent Experiments on Atomic Systems}, edited by F. S. Levin and D. A. Micha
(Plenum, New York, 1993)] and Drake
[Adv. At., Mol., Opt. Phys. {\bf 31}, 1 (1993)].
As a benchmark, we present a detailed tabulation of different energy levels.
\end{abstract}

 \maketitle

\section{Introduction}
The Rydberg states of few-electron atomic systems were investigated extensively from the mid-1980s to 1990s~\cite{DrakePlenum,Drake_Adv,Drake_Yan_PRA_1992,Drake_prl_1990,DrachmanHe,LundeenPrenum}.
According to the theory of Kelsey and Spruch~\cite{Kelsey,SpruchPrenum}, experimental and theoretical studies on high-$(n,L)$ states can test the Casimir-Polder effect, where $n$ and $L$ are, respectively, the principal and angular momentum quantum numbers of the Rydberg electron. The systems that have been studied include helium and lithium with one electron being excited to a high-$(n,L)$ state. A series of precision measurements were performed by Hessels {\it et al.}~\cite{hessels_He_1,hessels_He_2,hessels_He_3,hessels_He_4,hessels_He_5} on Rydberg states of helium using microwave spectroscopy. Hessels {\it et al.}~\cite{hessels_Li_1,hessels_Li_2} also did the radio-frequency measurements on lithium Rydberg states. On the theoretical side, a substantial work on Rydberg states of helium was
carried out independently by Drake~\cite{DrakePlenum,Drake_Adv,Drake_Yan_PRA_1992} and by Drachman~\cite{DrachmanHe} around the same period of time using the quantum mechanical perturbation method and the optical potential method, including relativistic and quantum electrodynamic (QED) effects. These methods are equivalent in nature and embody the picture of long-range interaction. A recent extension to higher angular momentum states of helium was done by El-Wazni and Drake~\cite{Wazni-Drake-pra-2009}.
Bhatia and Drachman~\cite{DrachmanLi1,DrachmanLi2,DrachmanLi3} also calculated relativistic and QED effects in the Rydberg states of lithium.  Later, Woods and Lundeen~\cite{Lundeen_Woods,Lundeen_Review} extended Drake and Drachman's work to more complex atoms, which allows for a high-$L$ Rydberg atom to have nonzero core angular momentum, for the purpose of modeling the effective potential and thus extracting core properties experimentally. Very recently, a new exotic Rydberg atom H$^{-+}$, which consists of
a Rydberg positron e$^+$ attached to the ground state H$^-$, was detected in the laboratory by Storry {\it et al.}~\cite{storry}. Since these Rydberg states are embedded in the Ps+H continuum, they are in fact resonant states~\cite{yan_ho}. It is therefore interesting to do theoretical calculation on these states and explore the spectrum of H$^{-+}$.

The main purpose of this paper is to present a complete calculation of nonrelativistic Rydberg energy levels using the standard perturbation method up to the order of $\langle x^{-10}\rangle$, where $x$ stands for the distance of the Rydberg particle relative to the core, and to compare our results with the work of Drake~\cite{Drake_Adv} and Drachman~\cite{DrachmanHe}. We find that there are several terms
of order $\langle x^{-10}\rangle$ missing in the work of Drake~\cite{Drake_Adv} and Drachman~\cite{DrachmanHe}. We also confirm a discrepancy that exists between Drake~\cite{Drake_Adv} and Drachman's~\cite{DrachmanHe} calculations. As a benchmark for future reference, we tabulate numerical values for the nonrelativistic energy levels of helium in various Rydberg states.

\section{Theory and Method}

\subsection{The Hamiltonian}

Consider an atomic or molecular system that consists of $n+2$ charged particles. The Hamiltonian of the system (in a.u.) is
\begin{eqnarray}
H &=& -\frac{1}{2m_0}\nabla_{{\bf R}_0}^2-\sum_{i=1}^n \frac{1}{2m_i}\nabla_{{\bf R}_i}^2
-\frac{1}{2m_{n+1}}\nabla_{{\bf R}_{n+1}}^2+\sum_{i> j\ge 0}^{n+1}\frac{q_i q_j}{|{\bf R}_i-{\bf R}_j|}\,,
\label{eq1}
\end{eqnarray}
where ${\bf R}_i$ is the position vector of the $i$th particle relative to the origin of a laboratory frame, with $0\le i \le n+1$, $m_i$ its mass, and $q_i$ its charge. We assume that the $(n+1)$th particle is far away from the "core", which is made up of the remaining $n+1$ particles. We also take the 0th particle as a reference one. In reality, it could be the nucleus. In order to eliminate the center of mass degree of freedom for the whole system, we make the
following coordinate transformations~\cite{yan_zhang_jpb}:
\begin{eqnarray}
{\bf X} &=&\frac{1}{M_T} \,{\sum_{j=0}^{n+1}m_j{\bf R}_j}\\
{\bf r}_i &=& {\bf R}_i-{\bf R}_0\,, \ \ i=1,2,\ldots, n\\
{\bf r}_{n+1}&=& {\bf R}_{n+1}-\frac{1}{M_C}\,\sum_{j=0}^n m_j {\bf R}_j\,,
\label{eq2}
\end{eqnarray}
where $M_T=\sum_{j=0}^{n+1}m_j$ is the total mass of the whole system, and $M_C=\sum_{j=0}^{n}m_j$ the total mass of the core. From the above expressions, we can see that ${\bf X}$ represents the position vector of the center of mass of the whole system, ${\bf r}_i$
is the position vector of $i$the particle in the core relative to the reference particle, and ${\bf r}_{n+1}$ is the position vector
of the Rydberg particle relative to the center of mass of the core. Thus, we have established a one to one transformation between the set $({\bf R}_0,{\bf R}_1,{\bf R}_2,\ldots,{\bf R}_{n},{\bf R}_{n+1})$ and the set $({\bf X},{\bf r}_1,{\bf r}_2,\ldots,{\bf r}_{n},{\bf r}_{n+1})$. The corresponding differential operators transform according to
\begin{eqnarray}
\nabla_{{\bf R}_0} &=& -\sum_{i=1}^n \nabla_i-\frac{m_0}{M_C}\nabla_{n+1}+\frac{m_0}{M_T}\nabla_{\bf X}\\
\nabla_{{\bf R}_i} &=& \nabla_i-\frac{m_i}{M_C}\nabla_{n+1}+\frac{m_i}{M_T}\nabla_{\bf X}\\
\nabla_{{\bf R}_{n+1}} &=& \nabla_{n+1}+\frac{m_{n+1}}{M_T}\nabla_{\bf X}\,,
\label{eq3}
\end{eqnarray}
where $\nabla_i\equiv\nabla_{{\bf r}_i}$ and $\nabla_{n+1}\equiv\nabla_{{\bf r}_{n+1}}$. After some simplification, the Hamiltonian (\ref{eq1}) can be rewritten in the form
\begin{eqnarray}
H &=& -\sum\limits_{i=1}^n\frac{1}{2\mu_i}\nabla_i^2-\frac{1}{2\mu_x}\nabla_{n+1}^2-\frac{1}{2M_T}\nabla_{\bf X}^2
-\frac{1}{m_0}\sum\limits_{i>j\ge 1}^n\nabla_i\cdot\nabla_j+\sum\limits_{i=1}^n\frac{q_iq_0}{r_i}+\sum\limits_{i>j\ge 1}^n\frac{q_iq_j}{r_{ij}}\nonumber\\
&&+\sum\limits_{i=1}^n \frac{q_iq_{n+1}}{\bigg|{\bf r}_i-{\bf r}_{n+1}-\frac{1}{M_C}\sum\limits_{j=1}^n m_j {\bf r}_j \bigg|}
+\frac{q_0q_{n+1}}{\bigg|{\bf r}_{n+1}+\frac{1}{M_C}\sum\limits_{j=1}^n m_j {\bf r}_j \bigg|}\,,
\label{eq4}
\end{eqnarray}
where ${\bf r}_{ij}={\bf r}_i-{\bf r}_j$ is the relative position between two core particles $i$ and $j$, $\mu_i=\frac{m_im_0}{m_i+m_0}$ ($1\le i\le n$) is the reduced mass of $i$th electron in the core with the reference particle 0, and $\mu_x=\frac{m_{n+1}M_C}{m_{n+1}+M_C}$ is the reduce mass of the Rydberg particle relative to the core. Since $H$ does not contain ${\bf X}$, ${\bf X}$ is a cyclic coordinate and thus can be ignored. Furthermore, the last two terms of (\ref{eq4}) may be combined by introducing
\begin{eqnarray}
\epsilon_{ij}&=& \delta_{ij}-m_j/M_C\,, \ \ 0\le i \le n\,, \ \ 1\le j \le n
\label{eq5}
\end{eqnarray}
{\it i.e.},
\begin{eqnarray}
&&\sum\limits_{i=1}^n \frac{q_iq_{n+1}}{\bigg|{\bf r}_i-{\bf r}_{n+1}-\frac{1}{M_C}\sum\limits_{j=1}^n m_j {\bf r}_j \bigg|}
+\frac{q_0q_{n+1}}{\bigg|{\bf r}_{n+1}+\frac{1}{M_C}\sum\limits_{j=1}^n m_j {\bf r}_j \bigg|}\nonumber \\
&=&\sum\limits_{i=1}^n \frac{q_iq_{n+1}}{\bigg|{\bf r}_{n+1}-\sum\limits_{j=1}^n \epsilon_{ij} {\bf r}_j \bigg|}
+\frac{q_0q_{n+1}}{\bigg|{\bf r}_{n+1}-\sum\limits_{j=1}^n \epsilon_{0j} {\bf r}_j \bigg|}\nonumber\\
&=&\sum\limits_{i=0}^n \frac{q_iq_{n+1}}{\bigg|{\bf r}_{n+1}-\sum\limits_{j=1}^n \epsilon_{ij} {\bf r}_j \bigg|}\,.
\label{eq6}
\end{eqnarray}
The Hamiltonian can thus be partitioned into the form
\begin{eqnarray}
H &=& H_c + H_x +V_{cx}, \ \ {\rm in} \ 2R_\infty\,,
\label{eq7}
\end{eqnarray}
where
\begin{eqnarray}
H_c &=& -\sum_{i=1}^n \frac{1}{2\mu_i}\nabla_i^2-\frac{1}{m_0}\sum_{i>j\ge 1}^n \nabla_i\cdot\nabla_j
+\sum_{i=1}^n\frac{q_0q_i}{r_i}+\sum_{r>j\ge 1}^n\frac{q_iq_j}{r_{ij}}\\
H_x &=& -\frac{1}{2\mu_x}\nabla_x^2+\frac{q_xq_c}{x}\\
V_{cx} &=& \sum_{i=0}^n \frac{q_iq_x}{\bigg| {\bf x}-\sum\limits_{j=1}^n\epsilon_{ij}{\bf r}_j   \bigg|}-\frac{q_cq_x}{x}
\label{eq8}
\end{eqnarray}
with $q_x\equiv q_{n+1}$, ${\bf x}\equiv {\bf r}_{n+1}$, and $q_c\equiv \sum_{j=0}^n q_j$ being the total charge of the core.
In (\ref{eq7}), $R_\infty$ is the Rydberg constant and $2R_\infty$ represents the atomic units of energy expressed in cm$^{-1}$.
It is clear that $H_c$ is the Hamiltonian of the core~\cite{yan_zhang_jpb}, $H_x$ the Hamiltonian of the Rydberg particle in the field of point charge $q_c$, and $V_{cx}$ the interaction potential energy between the core and the Rydberg particle.

For a highly excited Rydberg particle, we may assume that $|{\bf x}|> |\sum_{j=1}^n\epsilon_{ij}{\bf r}_j|$ for $0\le i\le n$. Under this condition, we have
\begin{eqnarray}
\frac{1}{|{\bf x}-{\bf d}|}&=&\sum_{\ell=0}^\infty\sum_{m=-\ell}^\ell\frac{4\pi}{2\ell+1}\frac{d^\ell}{x^{\ell+1}}
Y_{\ell m}^*(\hat{\bf x})Y_{\ell m}(\hat{\bf d})\,,
\label{eq9}
\end{eqnarray}
with ${\bf d}=\sum_{j=1}^n\epsilon_{ij}{\bf r}_j$. Using the formula~\cite{yan_zhang_jpb}
\begin{eqnarray}
Y_{\ell m}(\hat{{\bf r}}) &=& \sqrt{\frac{3}{4\pi}}\bigg(\prod_{s=1}^{\ell-1}\sqrt{\frac{2s+3}{s+1}}\bigg)
(\underbrace{\hat{\bf r}\otimes\hat{\bf r}\otimes\cdots\hat{\bf r}}_\ell)_m^{(\ell)}
\label{eq10}
\end{eqnarray}
with the understanding that $\prod_{s=1}^{\ell-1}\sqrt{\frac{2s+3}{s+1}}=1$ when $\ell=1$,
we obtain
\begin{eqnarray}
d^\ell Y_{\ell m}(\hat{\bf d}) &=& \sqrt{\frac{3}{4\pi}}
\bigg(\prod_{s=1}^{\ell-1} \sqrt{\frac{2s+3}{s+1}}\bigg)
(\underbrace{{\bf d}\otimes{\bf d}\otimes\cdots{\bf d}}_\ell)_m^{(\ell)}\nonumber\\
&=&\sqrt{\frac{3}{4\pi}}
\bigg(\prod_{s=1}^{\ell-1}\sqrt{\frac{2s+3}{s+1}}\bigg)
\sum_{j_1 j_2\cdots j_\ell \ge 1}^n (\epsilon_{ij_1}\epsilon_{ij_2}\cdots \epsilon_{ij_\ell})
({\bf r}_{j_1}\otimes{\bf r}_{j_2}\otimes\cdots{\bf r}_{j_\ell})_m^{(\ell)} \,.
\label{eq11}
\end{eqnarray}
Thus we have
\begin{eqnarray}
\sum_{i=0}^n\frac{q_iq_x}{\bigg|{\bf x}-\sum\limits_{j=1}^n\epsilon_{ij}{\bf r}_j\bigg|}
&=& \sum_{\ell=0}^\infty\sum_{m=-\ell}^\ell\frac{4\pi}{2\ell+1}[q_x x^{-\ell-1}
Y_{\ell m}^*(\hat{\bf x})]T_{\ell m}({\bf r}_1,{\bf r}_2,\ldots,{\bf r}_n)\,,
\label{eq12}
\end{eqnarray}
where
\begin{eqnarray}
&&T_{\ell m}({\bf r}_1,{\bf r}_2,\ldots,{\bf r}_n)
= \sqrt{\frac{3}{4\pi}}
\bigg(\prod_{s=1}^{\ell-1}\sqrt{\frac{2s+3}{s+1}}\bigg) \nonumber\\
&\times&
\sum_{j_1 j_2\cdots j_\ell \ge 1}^n \bigg(\sum_{i=0}^n q_i\epsilon_{ij_1}\epsilon_{ij_2}\cdots \epsilon_{ij_\ell}\bigg)
({\bf r}_{j_1}\otimes{\bf r}_{j_2}\otimes\cdots{\bf r}_{j_\ell})_m^{(\ell)}\,.
\label{eq13}
\end{eqnarray}
It is easy to see from (\ref{eq9}) that the term with $\ell=0$ is $1/x$ and its corresponding term in $V_{cx}$ is
$q_cq_x/x$, which cancels exactly with the second term in $V_{cx}$. In other words, there is no monopole contribution
to the interaction potential. Finally we obtain the following multipole expansion for the interaction potential energy $V_{cx}$,
where in each term the degree of freedom of the Rydberg particle is separated from the core coordinates
\begin{eqnarray}
V_{cx}&=& \sum_{\ell=1}^\infty\sum_{m=-\ell}^\ell\frac{4\pi}{2\ell+1}
\underbrace{[q_x x^{-\ell-1}Y_{\ell m}^*(\hat{\bf x})]}_{\rm Rydberg}\,
\underbrace{T_{\ell m}({\bf r}_1,{\bf r}_2,\ldots,{\bf r}_n)}_{\rm core}\,.
\label{eq14}
\end{eqnarray}

If we make the scaling transformation ${\bf x} \rightarrow \mu_x \,{\bf x}$, we obtain
the Hamiltonian
\begin{eqnarray}
H &=& h_c + h_x +v_{cx}, \ \ {\rm in} \ 2R_\infty\,,
\label{eq14x}
\end{eqnarray}
where
\begin{eqnarray}
h_c &=& -\sum_{i=1}^n \frac{1}{2\mu_i}\nabla_i^2-\frac{1}{m_0}\sum_{i>j\ge 1}^n \nabla_i\cdot\nabla_j
+\sum_{i=1}^n\frac{q_0q_i}{r_i}+\sum_{r>j\ge 1}^n\frac{q_iq_j}{r_{ij}}\\
h_x &=& \mu_x\bigg(-\frac{1}{2}\nabla_x^2+\frac{q_xq_c}{x}\bigg)\\
v_{cx} &=& \sum_{i=0}^n \frac{q_iq_x}{\bigg| {\bf \frac{1}{\mu_x}x}-\sum\limits_{j=1}^n\epsilon_{ij}{\bf r}_j   \bigg|}-\mu_x\frac{q_cq_x}{x}\nonumber\\
&=&\sum_{\ell=1}^\infty\sum_{m=-\ell}^\ell\frac{4\pi}{2\ell+1}\mu_x^{\ell+1}
[q_x x^{-\ell-1}Y_{\ell m}^*(\hat{\bf x})]\,
T_{\ell m}({\bf r}_1,{\bf r}_2,\ldots,{\bf r}_n)\,.
\label{eq14xx}
\end{eqnarray}

The above formulation is general for any system containing $n+2$ charged particles. If the system under consideration is
an atomic system with $n+1$ electrons and one nucleus, we assume that the 0th particle (the reference particle) is the nucleus with its mass $M$ and its nuclear charge $Z$. The Hamiltonian of the system becomes
\begin{eqnarray}
H &=& -\frac{1}{2\mu}\sum_{i=1}^n\nabla_i^2-\frac{1}{M}\sum_{i>j\ge 1}^n\nabla_i\cdot\nabla_j-\sum_{i=1}^n\frac{Z}{r_i}
+\sum_{i>j\ge 1}^n\frac{1}{r_{ij}}-\frac{1}{2\mu_x}\nabla_x^2
+\sum_{i=0}^n\frac{q_xq_i}{\bigg|{\bf x}-\sum\limits_{j=1}^n \epsilon_{ij}{\bf r}_j\bigg|}\,,
\label{eq15}
\end{eqnarray}
where $q_0=Z$, $q_i=-1$ ($1\le i\le n$), $q_x=-1$, $\mu$ is the reduced mass of the electron relative to the nucleus, and $\mu_x$
is the reduced mass of the Rydberg electron relative to the core mass $M+nm_e$. In order to see the finite nuclear mass effect more clearly, we make the following scaling transformations:
\begin{eqnarray}
{\bf r}_i &\rightarrow& \mu \,{\bf r}_i, \ \ i=1,2,\ldots n\\
{\bf x}   &\rightarrow& \mu_x \,{\bf x}\,.
\label{eq16}
\end{eqnarray}
The Hamiltonian (\ref{eq15}) can thus be transformed to
\begin{eqnarray}
H &=& h_c + h_x +v_{cx}, \ \ {\rm in} \ 2R_M\,,
\label{eq17}
\end{eqnarray}
where $R_M=\frac{\mu}{m_e}R_\infty$ and
\begin{eqnarray}
h_c &=& -\frac{1}{2}\sum_{i=1}^n \nabla_i^2-\frac{\mu}{M}\sum_{i>j\ge 1}^n \nabla_i\cdot\nabla_j
-\sum_{i=1}^n\frac{Z}{r_i}+\sum_{r>j\ge 1}^n\frac{1}{r_{ij}}\\
h_x &=& \frac{\mu_x}{\mu}\bigg(-\frac{1}{2}\nabla_x^2-\frac{Z-n}{x}\bigg)\\
v_{cx} &=& -\sum_{i=0}^n \frac{q_i}{\bigg| \frac{\mu}{\mu_x}{\bf x}-\sum\limits_{j=1}^n\epsilon_{ij}{\bf r}_j   \bigg|}+\frac{\mu_x}{\mu}\frac{(Z-n)}{x}\nonumber\\
&=&
\sum_{\ell=1}^\infty\sum_{m=-\ell}^\ell\frac{4\pi}{2\ell+1}\bigg(\frac{\mu_x}{\mu}\bigg)^{\ell+1}
[q_x x^{-\ell-1}Y_{\ell m}^*(\hat{\bf x})]\,
T_{\ell m}({\bf r}_1,{\bf r}_2,\ldots,{\bf r}_n)\,.
\label{eq18}
\end{eqnarray}

From now on, we use the following unified expressions for $h_x$ and $v_{cx}$
\begin{eqnarray}
h_x &=& a\,\bigg(-\frac{1}{2}\nabla_x^2-\frac{Z_1}{x}\bigg)\\
\label{eq18x_a1}
v_{cx} &=&
\sum_{\ell=1}^\infty\sum_{m=-\ell}^\ell C_\ell \,
u^*_{\ell m}({\bf x})\,
T_{\ell m}({\bf r}_1,{\bf r}_2,\ldots,{\bf r}_n)\,,
\label{eq18x}
\end{eqnarray}
where $a=\mu_x$ or $\mu_x/\mu$, $Z_1=-q_xq_c$ with $Z_1> 0$ in order to form a bound or quasi-bound Rydberg state,
\begin{eqnarray}
C_\ell &\equiv& \frac{4\pi}{2\ell+1}\,a^{\ell+1}\,q_x\,,
\label{eq18xx}
\end{eqnarray}
and
\begin{eqnarray}
u_{\ell m}({\bf x}) &\equiv & x^{-\ell-1}Y_{\ell m}(\hat{\bf x})
\label{eq37a}
\end{eqnarray}
denotes the irregular solid harmonics satisfying the Laplace equation $\nabla^2 u_{\ell m}({\bf x})=0$.
It should be mentioned that the Rydberg particle could be either an electron or positron, or any other charged particle.

\subsection{Perturbation expansion}

\subsubsection{Second-Order Energy: General Expression}

In (\ref{eq17}), we can treat $v_{cx}$ as a perturbation to the unperturbed Hamiltonian $H_0=h_c+h_x$, which is uncoupled.
The eigenvalue equations for $h_c$ and $h_x$ are respectively
\begin{eqnarray}
h_c \phi_{n_cL_cM_c} &=& \varepsilon_{n_c}(L_c)\phi_{n_cL_cM_c}\\
h_x\chi_{n_xL_xM_x} &=& e_{n_x}\chi_{n_xL_xM_x}\,,
\label{eq20}
\end{eqnarray}
where the eigenvalue $e_{n_x}$ only depends on the principal quantum number $n_x$ because of the
hydrogenic nature of $h_x$.
The initial eigenstates for $h_c$ and $h_x$ are assumed to be
\begin{eqnarray}
h_c \phi_0 &=& \varepsilon_0\phi_0 \\
h_x\chi_{n_0L_0M_0} &=& e_{n_0}\chi_{n_0L_0M_0}\,.
\label{eq21}
\end{eqnarray}
Thus
\begin{eqnarray}
H_0\Psi_0 &=& E_0\Psi_0
\label{eq21a}
\end{eqnarray}
where
\begin{eqnarray}
\Psi_0 &=& \phi_0\chi_{n_0L_0M_0}\\
E_0 &=&  \varepsilon_0+e_{n_0}\,.
\label{eq21b}
\end{eqnarray}

In this work, we only consider the case where $\phi_0$ is in an $S$ state, which results
in the consequence that the first-order energy correction due to $v_{cx}$ is zero, {\it i.e.},
\begin{eqnarray}
\Delta E_1 &=&\langle \Psi_0 |v_{cx}|\Psi_0\rangle = 0\,,
\label{eq22}
\end{eqnarray}
The reason why (\ref{eq22}) is valid is that there is no monopole term in the multipole expansion of $v_{cx}$ in (\ref{eq18x}).

The second-order energy correction can be calculated according to
\begin{eqnarray}
\Delta E_2 &=& \langle \Psi_0 | v_{cx} |\Psi_1\rangle\,,
\label{eq23a}
\end{eqnarray}
where
\begin{eqnarray}
|\Psi_1\rangle &=& \sum_n \frac{\langle \Psi_n |v_{cx}|\Psi_0\rangle}{E_0-E_n}|\Psi_n\rangle
\label{eq23}
\end{eqnarray}
and $n$ represents a set of quantum numbers describing an intermediate eigenstate of $H_0$, {\it i.e.},
\begin{eqnarray}
H_0\Psi_n &=& E_n\Psi_n
\label{eq24}
\end{eqnarray}
where
\begin{eqnarray}
\Psi_n &=& \phi_{n_cL_cM_c}\chi_{n_xL_xM_x}\\
E_n &=& \varepsilon_{n_c}(L_c)+e_{n_x}\,.
\label{eq25}
\end{eqnarray}
We first denote the excitation energies for the core and the Rydberg electron by
\begin{eqnarray}
\delta \varepsilon_{n_c}(L_c) &=& \varepsilon_{n_c}(L_c)-\varepsilon_0\\
\delta e_{n_x} &=& e_{n_x}-e_{n_0}\,.
\label{eq26}
\end{eqnarray}
Considering the Rydberg particle is in a highly excited state, we make the following key assumption that~\cite{Drake_Adv}
\begin{eqnarray}
|\delta e_{n_x}| &<  |\delta \varepsilon_{n_c}(L_c)|\,.
\label{eq27}
\end{eqnarray}
In the above we have implicitly assumed that $\delta \varepsilon_{n_c}(L_c)\ne 0$.
Now we can perform the following expansion
\begin{eqnarray}
\frac{1}{E_0-E_n}&=&
-\frac{1}{\delta \varepsilon_{n_c}(L_c)}\frac{1}{1+\frac{\delta e_{n_x}}{\delta \varepsilon_{n_c}(L_c)}}
=\sum_{i=0}^\infty (-1)^{i+1}\frac{(\delta e_{n_x})^i}{(\delta \varepsilon_{n_c}(L_c))^{i+1}}
\label{eq28}
\end{eqnarray}
Substituting (\ref{eq28}) into (\ref{eq23}) yields
\begin{eqnarray}
|\Psi_1\rangle &=& \sum_{i=0}^\infty (-1)^{i+1}\sum_{n_cL_cM_c}\sum_{n_xL_xM_x}\langle \phi_{n_cL_cM_c}\chi_{n_xL_xM_x}
|v_{cx}|\phi_0\chi_{n_0L_0M_0}\rangle\nonumber\\
&\times&\frac{h_s^i}{(\delta \varepsilon_{n_c}(L_c))^{i+1}}
|\phi_{n_cL_cM_c}\chi_{n_xL_xM_x}\rangle\,,
\label{eq29}
\end{eqnarray}
where we have applied the eigenvalue equation (\ref{eq20}) of $h_x$
\begin{eqnarray}
(\delta e_{n_x})^i |\chi_{n_xL_xM_x}\rangle &=&
h_s^i |\chi_{n_xL_xM_x}\rangle\,,
\label{eq30}
\end{eqnarray}
with the definition of $h_s\equiv h_x-e_{n_0}$ operating on the Rydberg electron. Now the second-order energy correction
(\ref{eq23a}) becomes
\begin{eqnarray}
\Delta E_2 &=& \sum_{i=0}^\infty (-1)^{i+1}\sum_{n_cL_cM_c}\sum_{n_xL_xM_x}\langle \phi_{n_cL_cM_c}\chi_{n_xL_xM_x}
|v_{cx}|\phi_0\chi_{n_0L_0M_0}\rangle\nonumber\\
&\times&\frac{1}{(\delta \varepsilon_{n_c}(L_c))^{i+1}}
\langle \phi_{0}\chi_{n_0L_0M_0}
|v_{cx}h_s^i|\phi_{n_cL_cM_c}\chi_{n_xL_xM_x}\rangle\,.
\label{eq31}
\end{eqnarray}
Substituting (\ref{eq18x}) into (\ref{eq31}) and using the Wigner-Eckart theorem for the matrix element $T_{\ell m}$
\begin{eqnarray}
\langle\phi_{n_cL_cM_c} |T_{\ell m}|\phi_0\rangle &=& (-1)^{L_c-M_c}
\left(
\begin{array}{ccc}
L_c & \ell & 0 \\
-M_c & m & 0 \\
\end{array}
\right)
\langle \phi_{n_cL_c} \|T_{\ell }\|\phi_0\rangle \nonumber\\
&=&\frac{1}{\sqrt{2L_c+1}}\delta_{\ell L_c}\delta_{mM_c}\langle \phi_{n_cL_c} \|T_{\ell }\|\phi_0\rangle\,,
\label{eq32}
\end{eqnarray}
we arrive at
\begin{eqnarray}
&&\langle \phi_{n_cL_cM_c}\chi_{n_xL_xM_x}|v_{cx}|\phi_0\chi_{n_0L_0M_0}\rangle \nonumber\\
&&=C_{L_c}\frac{1}{\sqrt{2L_c+1}}\langle \phi_{n_cL_c} \|T_{L_c}\|\phi_0\rangle
\langle \chi_{n_xL_xM_x}|u_{L_cM_c}^*({\bf x})|\chi_{n_0L_0M_0}\rangle\,,
\label{eq33}
\end{eqnarray}
where $C_{L_c}$ is defined in (\ref{eq18xx}).
It is noted here that $L_c\ge 1$ in (\ref{eq33}), as indicated in (\ref{eq18x}).
Similarly,
\begin{eqnarray}
&&\langle \phi_{0}\chi_{n_0L_0M_0}|v_{cx}h_s^i|\phi_{n_cL_cM_c}\chi_{n_xL_xM_x}\rangle \nonumber\\
&&=C_{L_c}(-1)^{L_c}\frac{1}{\sqrt{2L_c+1}}\langle \phi_{0} \|T_{L_c }\|\phi_{n_cL_c}\rangle
\langle \chi_{n_0L_0M_0}|u_{L_cM_c}({\bf x})h_s^i|\chi_{n_xL_xM_x}\rangle\,.
\label{eq35}
\end{eqnarray}
Substituting (\ref{eq33}) and (\ref{eq35}) into (\ref{eq31})
leads to the final expression for $\Delta E_2$
\begin{eqnarray}
\Delta E_2 &=& \sum_{i=0}^\infty (-1)^{i+1}\sum_{n_cL_c}C_{L_c}^2\frac{1}{2L_c+1}
\frac{|\langle \phi_{0} \|T_{L_c }\|\phi_{n_cL_c}\rangle|^2}{(\delta \varepsilon_{n_c}(L_c))^{i+1}}w^{(2)}_i(L_c)\,.
\label{eq36}
\end{eqnarray}
In the above $w^{(2)}_i(L_c)$ is the quantity that describes the Rydberg particle and is given by
\begin{eqnarray}
w^{(2)}_i(L_c) &=& \langle \chi_{n_0L_0M_0}|\mathcal{\hat{U}}_i(L_c)|\chi_{n_0L_0M_0}\rangle\,,
\label{eq37}
\end{eqnarray}
where the operator $\mathcal{\hat{U}}_i(\ell)$ is defined by
\begin{eqnarray}
\mathcal{\hat{U}}_i(\ell)&\equiv& \sum_m u_{\ell m}\,h_s^i\, u^*_{\ell m}\,.
\label{eq36a}
\end{eqnarray}
It is seen that $\mathcal{\hat{U}}_i(\ell)$ is an Hermitian operator.
In obtaining (\ref{eq36}), the following two relations have been used, namely, the closure relation
\begin{eqnarray}
\sum_{n_xL_xM_x}|\chi_{n_xL_xM_x}\rangle\langle \chi_{n_xL_xM_x}| &=& I
\label{eq38}
\end{eqnarray}
and
\begin{eqnarray}
\langle \phi_{n_cL_c} \|T_{L_c}\|\phi_0\rangle &=&
(-1)^{L_c}\langle \phi_{0} \|T_{L_c }\|\phi_{n_cL_c}\rangle^*\,.
\label{eq39}
\end{eqnarray}
It would be convenient to define the $2^{L_c}$-pole ``generalized polarizability" for the state of the $S$-symmetric core
\begin{eqnarray}
\alpha(i,L_c) &\equiv & \frac{2^{3-i}\pi}{(2L_c+1)^2}\sum_{n_c}
\frac{|\langle \phi_{0} \|T_{L_c }\|\phi_{n_cL_c}\rangle|^2}{(\delta \varepsilon_{n_c}(L_c))^{i+1}}\,.
\label{eq40}
\end{eqnarray}
In fact for the first few values of $i$, we have
\begin{eqnarray}
\alpha(0,L_c) &=& \alpha_{L_c}\\
\alpha(1,L_c) &=& \beta_{L_c}\\
\alpha(2,L_c) &=& \gamma_{L_c}\\
\alpha(3,L_c) &=& \delta_{L_c}\\
\alpha(4,L_c) &=& \varsigma_{L_c}\\
\alpha(5,L_c) &=& \eta_{L_c}\\
\alpha(6,L_c) &=& \theta_{L_c}\\
\alpha(7,L_c) &=& \iota_{L_c}
\label{eq41}
\end{eqnarray}
as defined by Drake~\cite{Drake_Adv} up to $i=3$. We therefore have the final expression for the second-order energy correction
\begin{eqnarray}
\Delta E_2 &=& \sum_{i=0}^\infty\sum_{L_c=1}^\infty\delta e_2(i,L_c)\,,
\label{eq42}
\end{eqnarray}
where
\begin{eqnarray}
\delta e_2(i,L_c) &=q_x^2& (-1)^{i+1}\frac{2^{i+1}\pi}{2L_c+1}a^{2L_c+2}\alpha(i,L_c)w^{(2)}_i(L_c)\,.
\label{eq43}
\end{eqnarray}

\subsubsection{Second-Order Energy: Calculations}

Consider $w^{(2)}_0(L_c)$ first. Using the formula
\begin{eqnarray}
\sum_{m=-\ell}^{\ell}Y_{\ell m}(\hat{\bf x})Y^*_{\ell m}(\hat{\bf x}) &=& \frac{2\ell+1}{4\pi}\,,
\label{eq44x}
\end{eqnarray}
we have
\begin{eqnarray}
w^{(2)}_0(L_c) &=& \frac{2L_c+1}{4\pi}\langle\chi_{n_0L_0M_0}|x^{-2L_c-2}|\chi_{n_0L_0M_0}\rangle
=\frac{2L_c+1}{4\pi}\langle x^{-2L_c-2}\rangle_{n_0L_0}
\label{eq45x}
\end{eqnarray}
with $| \ \rangle_{n_0L_0}\equiv |\chi_{n_0L_0M_0}\rangle$.
It should be pointed out that $\langle x^{-s}\rangle_{n_0L_0}$ diverges
unless $s\le 2L_0+2$. The analytical expressions for $\langle x^{-s}\rangle_{n_0L_0}$ with $s$ up to 16 are given explicitly by
Drake and Swainson~\cite{drake_swainson_hydrogen}. Thus the result for $i=0$ is
\begin{eqnarray}
\delta e_2(0,L_c) &=& -\frac{1}{2}q_x^2\,a^{2L_c+2}\alpha_{L_c}\langle x^{-2L_c-2}\rangle_{n_0L_0}\,.
\label{eq46x}
\end{eqnarray}

For the case of $i=1$, we first notice that
\begin{eqnarray}
h_s |\chi_{n_0L_0M_0}\rangle &=& 0\,.
\label{eq47x}
\end{eqnarray}
Thus we have
\begin{eqnarray}
&&\sum_{M_c={-L_c}}^{L_c}\langle \chi_{n_0L_0M_0}|[u_{L_cM_c},[h_s,u^*_{L_cM_c}]]|\chi_{n_0L_0M_0}\rangle \nonumber\\
&=&\sum_{M_c={-L_c}}^{L_c}\langle \chi_{n_0L_0M_0}|u_{L_cM_c}h_su^*_{L_cM_c}|\chi_{n_0L_0M_0}\rangle
+\sum_{M_c={-L_c}}^{L_c}\langle \chi_{n_0L_0M_0}|u^*_{L_cM_c}h_su_{L_cM_c}|\chi_{n_0L_0M_0}\rangle\nonumber\\
&=&2\sum_{M_c={-L_c}}^{L_c}\langle \chi_{n_0L_0M_0}|u_{L_cM_c}h_su^*_{L_cM_c}|\chi_{n_0L_0M_0}\rangle\,,
\label{eq48x}
\end{eqnarray}
where we have used the property that $u^*_{L_cM_c}=(-1)^{M_c}u_{L_c-M_c}$, as well as the fact that any summation above will be the same when switching $M_c$ to $-M_c$. Therefore, the $w^{(2)}_1(L_c)$ can be recast into
\begin{eqnarray}
w^{(2)}_1(L_c) &=&\frac{1}{2}\sum_{M_c}\langle \chi_{n_0L_0M_0}|[u_{L_cM_c},[h_s,u^*_{L_cM_c}]]|\chi_{n_0L_0M_0}\rangle\nonumber\\
&=&-\frac{a}{4}\sum_{M_c}\langle \chi_{n_0L_0M_0}|[u_{L_cM_c},[\nabla^2,u^*_{L_cM_c}]]|\chi_{n_0L_0M_0}\rangle\,,
\label{eq49x}
\end{eqnarray}
where we have ignored the subscript $x$ in $\nabla^2$.
Since $u_{\ell m}({\bf x})$ is a harmonic function, it satisfies the Laplace equation $\nabla^2u_{\ell m}=0$.
It is therefore straightforward to show the following operator relations
\begin{eqnarray}
&&[\nabla^2, u_{L_cM_c}] = 2(\nabla u_{L_cM_c})\cdot\nabla
\label{eq50a}\\
&&[u_{L_cM_c},[\nabla^2,u^*_{L_cM_c}]] = -2 (\nabla u_{L_cM_c})(\nabla u^*_{L_cM_c})\,.
\label{eq50x}
\end{eqnarray}
Furthermore, using the following two formulas~\cite{varshalovich}
\begin{eqnarray}
\nabla u_{\ell m}({\bf x})&=& \sqrt{(\ell+1)(2\ell+1)}|{\bf x}|^{-\ell-2}{\bf Y}_{\ell\ell+1m}(\hat{\bf x})
\label{eq51x}
\end{eqnarray}
and
\begin{eqnarray}
\sum_{M=-J}^J {\bf Y}_{J\ell M}(\hat{\bf x})\cdot {\bf Y}^*_{J\ell M}(\hat{\bf x}) &=& \frac{2J+1}{4\pi}\,,
\label{eq52}
\end{eqnarray}
where ${\bf Y}_{J\ell M}(\hat{\bf x})$ is the vector spherical harmonics, we arrive at
\begin{eqnarray}
w^{(2)}_1(L_c) &=& \frac{a}{8\pi}(L_c+1)(2L_c+1)^2\langle x^{-2L_c-4}  \rangle_{n_0L_0} \,.
\label{eq53x}
\end{eqnarray}
Finally the corresponding energy correction for given $i=1$ and $L_c$ is
\begin{eqnarray}
\delta e_2(1,L_c) &=& \frac{1}{2}q_x^2\,a^{2L_c+3}(L_c+1)(2L_c+1)\beta_{L_c}\langle x^{-2L_c-4}  \rangle_{n_0L_0}\,.
\label{eq54x}
\end{eqnarray}

Now we consider the general case where $i$ is an arbitrary positive integer. We first consider the following expression
\begin{eqnarray}
h_s[f(x)Y_{\ell m}(\hat {\bf x})]\,.
\label{eq43a}
\end{eqnarray}
Noting that
\begin{eqnarray}
h_s &=& -\frac{a}{2}\nabla^2-\frac{aZ_1}{x}-e_{n_0}\nonumber\\
&=& -\frac{a}{2}\bigg[-\frac{L^2}{x^2}+\frac{1}{x^2}\frac{\partial}{\partial x}\bigg(x^2\frac{\partial}{\partial x}\bigg)\bigg]
-\frac{aZ_1}{x}-e_{n_0}\,,
\label{eq44}
\end{eqnarray}
where $L^2$ is the angular momentum squared, we arrive at
\begin{eqnarray}
h_s [f(x)Y_{\ell m}(\hat {\bf x})] &=& [h_r(\ell)f(x)] Y_{\ell m}(\hat {\bf x})\,.
\label{eq45}
\end{eqnarray}
In the above, $h_r(\ell)$ is defined by
\begin{eqnarray}
h_r(\ell) &\equiv& \frac{a}{2}\frac{\ell (\ell+1)}{x^2}-\frac{a}{2}\frac{1}{x^2}\frac{d}{dx}\bigg(x^2\frac{d}{dx}\bigg)-\frac{aZ_1}{x}-e_{n_0}\,,
\label{eq46}
\end{eqnarray}
acting only on the radial function $f(x)$. The repeated use of (\ref{eq45}) yields
\begin{eqnarray}
h^p_s [f(x)Y_{\ell m}(\hat {\bf x})] &=& [h^p_r (\ell) f(x)]Y_{\ell m}(\hat {\bf x})\,.
\label{eq47}
\end{eqnarray}
It is seen that the operator $h_s$, when applying to $f(x)Y_{\ell m}(\hat {\bf x})$, only changes the radial part,
not the angular part.

Consider $w^{(2)}_i(L_c)$.
Let the wave function of the Rydberg electron be
\begin{eqnarray}
|\chi_{n_0L_0M_0}\rangle &=& R_{n_0L_0}(x)Y_{L_0 M_0}(\hat{\bf x})\,.
\label{eq48}
\end{eqnarray}
Then we have
\begin{eqnarray}
w^{(2)}_i(L_c)
&=& \sum_{M_c}\int x^2 \,dx \,d\Omega\, R_{n_0L_0}(x)Y^*_{L_0M_0}(\hat{\bf x})
x^{-L_c-1}Y_{L_cM_c}(\hat{\bf x})\nonumber\\
&\times& h_s^i x^{-L_c-1}Y^*_{L_cM_c}(\hat{\bf x})R_{n_0L_0}(x)Y_{L_0 M_0}(\hat{\bf x})\,.
\label{eq49}
\end{eqnarray}
Note that
{\small
\begin{eqnarray}
Y^*_{L_c M_c}(\hat{\bf x})Y_{L_0 M_0}(\hat{\bf x})&=&
(-1)^{M_c}\sum_{\Omega_1 \omega_1}\frac{(L_c,L_0,\Omega_1)^{1/2}}{\sqrt{4\pi}}
\left(
\begin{array}{ccc}
L_c & L_0 & \Omega_1 \\
0 & 0 & 0 \\
\end{array}
\right)
\left(
\begin{array}{ccc}
L_c & L_0 & \Omega_1 \\
-M_c & M_0 & \omega_1 \\
\end{array}
\right)
Y^*_{\Omega_1 \omega_1}(\hat{\bf x})\,,
\label{eq50}
\end{eqnarray}
}
where the notation $(\ell_1,\ell_2,\ldots)\equiv (2\ell_1+1)(2\ell_2+1)\ldots$,
and
{\small
\begin{eqnarray}
\int d\Omega
Y^*_{L_0 M_0}(\hat{\bf x})Y_{L_c M_c}(\hat{\bf x}) Y^*_{\Omega_1 \omega_1}(\hat{\bf x})&=&
(-1)^{M_c}\frac{(L_0,\Omega_1,L_c)^{1/2}}{\sqrt{4\pi}}
\left(
\begin{array}{ccc}
L_0 & \Omega_1 & L_c \\
0 & 0 & 0 \\
\end{array}
\right)
\left(
\begin{array}{ccc}
L_0 & \Omega_1 & L_c \\
M_0 & \omega_1 & -M_c \\
\end{array}
\right)\,.
\label{eq51}
\end{eqnarray}
}
The sum over $M_c$ and $\omega_1$ in $w_i^{(2)}(L_c)$ can then be performed according to
 \begin{eqnarray}
 \sum_{M_c\omega_1}
 \left(
\begin{array}{ccc}
L_c & L_0 & \Omega_1 \\
-M_c & M_0 & \omega_1 \\
\end{array}
\right)
\left(
\begin{array}{ccc}
L_0 & \Omega_1 & L_c \\
M_0 & \omega_1 & -M_c \\
\end{array}
\right)
&=&\frac{1}{2L_0+1}\,.
\label{eq52}
\end{eqnarray}
With all these above, we finally have
\begin{eqnarray}
w_i^{(2)}(L_c) &=& \frac{2L_c+1}{4\pi}\sum_{\Omega_1}(2\,\Omega_1+1)
\left(
\begin{array}{ccc}
L_c & L_0 & \Omega_1 \\
0 & 0 & 0 \\
\end{array}
\right)^2\nonumber\\
&\times&\int_0^\infty dx\, x^{-L_c+1}R_{n_0L_0}(x)\{h_r^i(\Omega_1)[x^{-L_c-1}R_{n_0L_0}(x)]\}\,.
\label{eq53}
\end{eqnarray}

In (\ref{eq53}) after the application of $h_r^i(\Omega_1)$ on $x^{-L_c-1}R_{n_0L_0}(x)$, we need to evaluate the following type of integral:
\begin{eqnarray}
J(s,n) &=& \int_0^\infty dx\, x^{-s} R_{n_0L_0}(x)R^{(n)}_{n_0L_0}(x)\,,
\label{eq90}
\end{eqnarray}
where $s$ is a positive integer and $ R^{(n)}_{n_0L_0}(x)$ denotes the $n$th-order derivative of
$ R_{n_0L_0}(x)$. We start by applying the Hamiltonian (\ref{eq44}) to the wave function of the Rydberg electron $R_{n_0L_0}(x)Y_{L_0 M_0}(\hat{\bf x})$, resulting in the following equation:
\begin{eqnarray}
\frac{2}{x}R'_{n_0L_0}(x)+R''_{n_0L_0}(x)-\frac{L_0(L_0+1)}{x^2}R+\frac{2Z_1}{x}R+\frac{2e_{n_0}}{a}R &=&0\,.
\label{eq91}
\end{eqnarray}
Performing $\frac{d^n}{dx^n}$ on the above equation, expanding the derivatives of products by using the Leibniz formula, and finally integrating $\int_0^\infty dx\, x^{-s}R_{n_0L_0}(x)\cdots$ throughout, we arrive at the recursion relation
\begin{eqnarray}
&&J(s,n+2)+2\sum_{i=0}^n(-1)^{n-i}\,\frac{n!}{i!}J(n-i+1+s,i+1)\nonumber\\
&&-L_0(L_0+1)\sum_{i=0}^n(-1)^{n-i}\,\frac{n!(n-i+1)}{i!}J(n-i+2+s,i)\nonumber\\
&&+2Z_1\sum_{i=0}^n(-1)^{n-i}\,\frac{n!}{i!}J(n-i+1+s,i)+\frac{2e_{n_0}}{a}J(s,n)=0\,.
\label{eq92}
\end{eqnarray}
The above recursion relation shows that, in order to calculate $J(s,n)$, one needs to know $J(s',m)$ with $0\le m\le n-1$. The initial integrals are
\begin{eqnarray}
J(s,0) &=&\langle x^{-s-2} \rangle_{n_0L_0}\,,
\label{eq93}
\end{eqnarray}
and $J(s,1)$ that can be evaluated as follows.
\begin{eqnarray}
&&J(s,1) = \frac{1}{2}\int_0^\infty \,x^{-s}\,dR^2_{n_0L_0}(x)\nonumber\\
&&=\frac{1}{2}x^{-s}R^2_{n_0L_0}(x)|_0^\infty+\frac{1}{2}s\int_0^\infty dx\, x^{-s-1}R^2_{n_0L_0}(x)\nonumber\\
&&=\frac{s}{2}\langle x^{-s-3} \rangle_{n_0L_0}\,.
\label{eq94}
\end{eqnarray}
In the above, the surface term vanishes at $\infty$ because $R_{n_0L_0}(x)$ decays to zero exponentially; it also vanishes at $x=0$, provided $L_0>s/2$ due to the fact that $R_{n_0L_0}(x)\sim x^{L_0}$ as $x\sim 0$.

It is advantageous to transform $e^j_{n_0} \langle x^{-s} \rangle_{n_0L_0}$ into a series of $\langle x^{-s'} \rangle_{n_0L_0}$.
This can be done by using the so-called hypervirial theorem~\cite{DrachmanHe}:
\begin{eqnarray}
\langle \chi_{n_0L_0M_0}|[x^p\frac{d}{dx},h_x]|\chi_{n_0L_0M_0}\rangle &=&0\,,
\label{eq95}
\end{eqnarray}
where
\begin{eqnarray}
h_x &=& -\frac{a}{2}\bigg(\frac{2}{x}\frac{d}{dx}+\frac{d^2}{dx^2}\bigg)+\frac{a}{2}\frac{L_0(L_0+1)}{x^2}-\frac{aZ_1}{x}\,.
\label{eq96}
\end{eqnarray}
We note that
\begin{eqnarray}
[x^p\frac{d}{dx},h_x]&=&[x^p,h_x]\frac{d}{dx}+x^p[\frac{d}{dx},h_x]\,.
\label{eq97}
\end{eqnarray}
It is a straightforward matter to find that
\begin{eqnarray}
&&[x^p,h_x] = \frac{a}{2}p(p+1)x^{p-2}+apx^{p-1}\frac{d}{dx}\\
&&[\frac{d}{dx},h_x] = a\frac{1}{x^2}\frac{d}{dx}-aL_0(L_0+1)\frac{1}{x^3}+aZ_1\frac{1}{x^2}\,.
\label{eq98}
\end{eqnarray}
Substituting the above into (\ref{eq97}), the hypervirial theorem (\ref{eq95}) reads
\begin{eqnarray}
&&p\langle x^{p-1}\frac{d^2}{dx^2}\rangle_{n_0L_0}
+[1+\frac{1}{2}p(p+1)]\langle x^{p-2}\frac{d}{dx} \rangle_{n_0L_0}\nonumber\\
&&-L_0(L_0+1)\langle x^{p-3} \rangle_{n_0L_0}+Z_1\langle x^{p-2} \rangle_{n_0L_0}=0\,.
\label{eq99}
\end{eqnarray}
The second-order derivative operator above can be replaced by
\begin{eqnarray}
\frac{d^2}{dx^2}&=&-\frac{2}{x}\frac{d}{dx}+L_0(L_0+1)\frac{1}{x^2}-\frac{2Z_1}{x}-\frac{2e_{n_0}}{a}\,.
\label{eq100}
\end{eqnarray}
After putting it back into (\ref{eq99}) and then using $\langle x^{p-2}d/dx \rangle_{n_0L_0}=-\frac{p}{2}\langle x^{p-3} \rangle_{n_0L_0}$ from (\ref{eq94}), one finally arrives at
\begin{eqnarray}
e_{n_0}\langle x^{p-1} \rangle_{n_0L_0} &=&a\frac{1-2p}{2p}Z_1\langle x^{p-2}\rangle_{n_0L_0}
+a\frac{p-1}{2p}[L_0(L_0+1)-\frac{1}{4}p(p-2)]\langle x^{p-3} \rangle_{n_0L_0}\,.
\label{eq101}
\end{eqnarray}
The term $e^j_{n_0} \langle x^{-s} \rangle_{n_0L_0}$ can be calculated by repeated use of (\ref{eq101}).

With the above preparations, we are now in a position to evaluate $w^{(2)}_i(L_c)$ and then the second-order energy corrections $\delta e_2(i,L_c)$, with the help of software Maple. We have already obtained $w^{(2)}_0(L_c)$ and $w^{(2)}_1(L_c)$ in (\ref{eq45x}) and (\ref{eq53x}) respectively.
For $w^{(2)}_2(L_c)$ we have
\begin{eqnarray}
w^{(2)}_2(L_c)&=&-\frac{Z_1a^2(L_c+1)^2(2L_c+1)}{4\pi (2L_c+3)}\langle x^{-2L_c-5}\rangle_{n_0L_0}\nonumber\\
&+&\frac{a^2(L_c+1)^2(L_c+2)(2L_c+1)^2}{8\pi}\bigg[1+\frac{L_0(L_0+1)}{(L_c+1)(2L_c+3)}\bigg]\langle x^{-2L_c-6}\rangle_{n_0L_0}\,.
\label{eq101x}
\end{eqnarray}
For $w^{(2)}_3(L_c)$ we have
\begin{eqnarray}
w^{(2)}_3(L_c)&=&-\frac{Z_1a^3(L_c+1)^2(L_c+2)(2L_c+1)(6L_c+11)}{8\pi (2L_c+5)}\langle x^{-2L_c-7}\rangle_{n_0L_0}\nonumber\\
&+&\frac{a^3(L_c+1)^2(L_c+2)(L_c+3)(2L_c+1)^2(2L_c+3)}{16\pi}\nonumber\\
&\times&\bigg[1+\frac{3L_0(L_0+1)}{(L_c+1)(2L_c+5)}\bigg]\langle x^{-2L_c-8}\rangle_{n_0L_0}\,.
\label{eq101xx}
\end{eqnarray}

In the following, we list some special values of the second-order energy corrections. For $\delta e_2(2,L_c)$ we have
\begin{eqnarray}
\delta e_2(2,1) &=& q_x^2\,a^6\,\gamma_1\bigg\{\frac{8Z_1}{5}\langle x^{-7}\rangle_{n_0L_0}
-36\bigg(1+\frac{L_0(L_0+1)}{10}\bigg)\langle x^{-8}\rangle_{n_0L_0}\bigg\}\,,
\label{eq102}
\end{eqnarray}
\begin{eqnarray}
\delta e_2(2,2) &=& q_x^2\,a^8\,\gamma_2\bigg\{\frac{18Z_1}{7}\langle x^{-9}\rangle_{n_0L_0}
-180\bigg(1+\frac{L_0(L_0+1)}{21}\bigg)\langle x^{-10}\rangle_{n_0L_0}\bigg\}\,,
\label{eq103}
\end{eqnarray}
\begin{eqnarray}
\delta e_2(2,3) &=& q_x^2\,a^{10}\,\gamma_3\bigg\{\frac{32Z_1}{9}\langle x^{-11}\rangle_{n_0L_0}
-560\bigg(1+\frac{L_0(L_0+1)}{36}\bigg)\langle x^{-12}\rangle_{n_0L_0}\bigg\}\,.
\label{eq104}
\end{eqnarray}
%In general, we have
%\begin{eqnarray}
%&&\delta e_2(2,L_c) = a^{2L_c+4}\,\gamma_{L_c}(L_c+1)^2\bigg\{\frac{2Z_1}{2L_c+3}\langle x^{-2L_c-5}\rangle_{n_0L_0}\nonumber\\
%&&
%-(L_c+2)(2L_c+1)\bigg(1+\frac{L_0(L_0+1)}{(L_c+1)(2L_c+3)}\bigg)\langle x^{-2L_c-6}\rangle_{n_0L_0}\bigg\}\,,
%\label{eq105}
%\end{eqnarray}
%which has been verified extensively by Maple.
For $\delta e_2(3,L_c)$, we have
\begin{eqnarray}
\delta e_2(3,1) &=& q_x^2\,a^{7}\,\delta_1\bigg\{-\frac{408Z_1}{7}\langle x^{-9}\rangle_{n_0L_0}
+720\bigg(1+\frac{3}{14}L_0(L_0+1)\bigg)\langle x^{-10}\rangle_{n_0L_0}\bigg\}\,,
\label{eq106}
\end{eqnarray}
\begin{eqnarray}
\delta e_2(3,2) &=& q_x^2\,a^{9}\,\delta_2\bigg\{-184Z_1\langle x^{-11}\rangle_{n_0L_0}
+6300\bigg(1+\frac{1}{9}L_0(L_0+1)\bigg)\langle x^{-12}\rangle_{n_0L_0}\bigg\}\,,
\label{eq107}
\end{eqnarray}
\begin{eqnarray}
\delta e_2(3,3) &=& q_x^2\,a^{11}\,\delta_3\bigg\{-\frac{4640Z_1}{11}\langle x^{-13}\rangle_{n_0L_0}
+30240\bigg(1+\frac{3}{44}L_0(L_0+1)\bigg)\langle x^{-14}\rangle_{n_0L_0}\bigg\}\,.
\label{eq108}
\end{eqnarray}
%Similarly we can obtain the following general expression:
%\begin{eqnarray}
%&&\delta e_2(3,L_c) = a^{2L_c+5}\,\delta_{L_c}(L_c+1)^2(L_c+2)\bigg\{-\frac{2Z_1(6L_c+11)}{2L_c+5}\langle %x^{-2L_c-7}\rangle_{n_0L_0}\nonumber\\
%&&
%+(L_c+3)(2L_c+1)(2L_c+3)\bigg(1+\frac{3}{(L_c+1)(2L_c+5)}L_0(L_0+1)\bigg)\langle x^{-2L_c-8}\rangle_{n_0L_0}\bigg\}\,.
%\label{eq109}
%\end{eqnarray}
For $\delta e_2(4,L_c)$, we have
\begin{eqnarray}
\delta e_2(4,1) &=& q_x^2\,a^{8}\,\varsigma_1\bigg\{-\frac{164Z_1^2}{7}\langle x^{-10}\rangle_{n_0L_0}
+\frac{16368Z_1}{7}\bigg(1+\frac{59}{1364}L_0(L_0+1)\bigg)\langle x^{-11}\rangle_{n_0L_0}\nonumber\\
&-&\frac{600}{7}(252+82L_0+83L_0^2+2L_0^3+L_0^4)\langle x^{-12}\rangle_{n_0L_0}\bigg\}\,,
\label{eq110}
\end{eqnarray}
\begin{eqnarray}
\delta e_2(4,2) &=& q_x^2\,a^{10}\,\varsigma_2\bigg\{-\frac{264Z_1^2}{5}\langle x^{-12}\rangle_{n_0L_0}
+\frac{140736Z_1}{11}\bigg(1+\frac{97}{3665}L_0(L_0+1)\bigg)\langle x^{-13}\rangle_{n_0L_0}\nonumber\\
&-&\frac{4200}{11}(792+142L_0+143L_0^2+2L_0^3+L_0^4)\langle x^{-14}\rangle_{n_0L_0}\bigg\}\,,
\label{eq111}
\end{eqnarray}
\begin{eqnarray}
\delta e_2(4,3) &=& q_x^2\,a^{12}\,\varsigma_3\bigg\{-\frac{3104Z_1^2}{33}\langle x^{-14}\rangle_{n_0L_0}
+\frac{6449600Z_1}{143}\bigg(1+\frac{427}{24186}L_0(L_0+1)\bigg)\langle x^{-15}\rangle_{n_0L_0}\nonumber\\
&-&\frac{158760}{143}\bigg(\frac{5720}{3}+218L_0+219L_0^2+2L_0^3+L_0^4\bigg)\langle x^{-16}\rangle_{n_0L_0}\bigg\}\,.
\label{eq112}
\end{eqnarray}
For $\delta e_2(5,L_c)$, we have
\begin{eqnarray}
\delta e_2(5,1) &=& q_x^2\,a^{9}\,\eta_1\bigg\{\frac{12096Z_1^2}{5}\langle x^{-12}\rangle_{n_0L_0}
-\frac{1283904Z_1}{11}\bigg(1+\frac{382}{3715}L_0(L_0+1)\bigg)\langle x^{-13}\rangle_{n_0L_0}\nonumber\\
&+&\frac{126000}{11}\bigg(\frac{396}{5}+34L_0+35L_0^2+2L_0^3+L_0^4\bigg)\langle x^{-14}\rangle_{n_0L_0}\bigg\}\,,
\label{eq113}
\end{eqnarray}
\begin{eqnarray}
\delta e_2(5,2) &=& q_x^2\,a^{11}\,\eta_2\bigg\{\frac{101712Z_1^2}{11}\langle x^{-14}\rangle_{n_0L_0}
-\frac{145356480Z_1}{143}\bigg(1+\frac{19889}{302826}L_0(L_0+1)\bigg)\langle x^{-15}\rangle_{n_0L_0}\nonumber\\
&+&\frac{11907000}{143}\bigg(\frac{1144}{5}+\frac{170}{3}L_0+\frac{173}{3}L_0^2+2L_0^3+L_0^4\bigg)\langle x^{-16}\rangle_{n_0L_0}\bigg\}\,,
\label{eq114}
\end{eqnarray}
For $\delta e_2(6,L_c)$, we have
\begin{eqnarray}
&&\delta e_2(6,1) = q_x^2\,a^{10}\,\theta_1\bigg\{\frac{42112Z_1^3}{55}\langle x^{-13}\rangle_{n_0L_0}
-\frac{12166784Z_1^2}{55}\bigg(1+\frac{2076}{95053}L_0(L_0+1)\bigg)\langle x^{-14}\rangle_{n_0L_0}\nonumber\\
&&+\frac{5824128Z_1}{715}\bigg(\frac{13632820}{15167}+\frac{6876328}{45501}L_0+\frac{6921829}{45501}L_0^2+2L_0^3+L_0^4\bigg)\langle x^{-15}\rangle_{n_0L_0}\nonumber\\
&&-\frac{588000}{143}\bigg(\frac{61776}{5}+6492L_0+6808L_0^2+633L_0^3+319L_0^4+3L_0^5+L_0^6\bigg)\langle x^{-16}\rangle_{n_0L_0}\bigg\}\,.
\label{eq116}
\end{eqnarray}

\subsubsection{Third-Order Energy}

The third-order energy correction can be calculated according to
\begin{eqnarray}
\Delta E_3 &=& \langle \Psi_1 | v_{cx} |\Psi_1\rangle\,,
\label{eq122}
\end{eqnarray}
where $\Psi_1$ is defined in (\ref{eq23}) and further expanded in (\ref{eq29}). In the above we have used the fact that $\Delta E_1=0$ (see (\ref{eq22})).
Using the similar procedure towards (\ref{eq36}) leads to the final expression for
$\Delta E_3$:
\begin{eqnarray}
\Delta E_3 &=& \sum_{i=0}^\infty\sum_{j=0}^\infty (-1)^{i+j}\sum_{L_c'\ell L_c\ge 1}
\frac{C_{L_c'}C_\ell C_{L_c}}{\sqrt{(L_c',L_c)}}w^{(3)}_c(i,j;L_c',\ell,L_c)w^{(3)}_{ij}(L_c',\ell,L_c)\,,
\label{eq123}
\end{eqnarray}
where the quantity $w^{(3)}_c(i,j;L_c',\ell,L_c)$ describing the core is defined by
\begin{eqnarray}
w^{(3)}_c(i,j;L_c',\ell,L_c) &\equiv &
\sum_{n_cn_c'}\frac
{\langle \phi_0\|T_{L_c'}\|\phi_{n_c'L_c'}\rangle \langle \phi_{n_c'L_c'} \|T_\ell\|\phi_{n_cL_c}\rangle
\langle \phi_{n_cL_c} \|T_{L_c}\|\phi_0\rangle  }
{[\delta \varepsilon_{n_c}(L_c)]^{i+1}[\delta \varepsilon_{n_c'}(L_c')]^{j+1}}\,,
\label{eq124}
\end{eqnarray}
and the quantity relevant to the Rydberg electron $w^{(3)}_{ij}(L_c',\ell,L_c)$ is defined by
\begin{eqnarray}
{\small
w^{(3)}_{ij}(L_c',\ell,L_c) \equiv \sum_{M_c' m M_c}(-1)^{M_c'}
\left(
\begin{array}{ccc}
L_c' & \ell & L_c \\
-M_c' & m & M_c \\
\end{array}
\right)
\langle \chi_{n_0L_0M_0}|u_{L_c'M_c'}\,h_s^j \,u^*_{\ell m}\,h_s^i \,u^*_{L_cM_c} |\chi_{n_0L_0M_0}\rangle\,.
}
\label{eq125}
\end{eqnarray}
The above expression can be simplified by integrating over the angular coordinates. Since
\begin{eqnarray}
&&w^{(3)}_{ij}(L_c',\ell,L_c) =
\sum_{M_c' m M_c}(-1)^{M_c'}
\left(
\begin{array}{ccc}
L_c' & \ell & L_c \\
-M_c' & m & M_c \\
\end{array}
\right)
\int x^2 \,dx \,d\Omega\, R_{n_0L_0}(x)Y^*_{L_0M_0}(\hat{\bf x})
x^{-L_c'-1}Y_{L_c'M_c'}(\hat{\bf x})\nonumber\\
&&\times h_s^j\, x^{-\ell-1}Y^*_{\ell m}(\hat{\bf x})\,h_s^i\,
x^{-L_c-1}Y^*_{L_c M_c}(\hat{\bf x})R_{n_0L_0}(x)Y_{L_0 M_0}(\hat{\bf x})\,,
\label{eq126}
\end{eqnarray}
the product of the two spherical harmonic functions $Y^*_{L_c M_c}(\hat{\bf x})$ and
$Y_{L_0 M_0}(\hat{\bf x})$ can be reduced to a single one $Y^*_{\Omega_1 \omega_1}(\hat{\bf x})$ according to
(\ref{eq50}). Then using (\ref{eq47}) one obtains
\begin{eqnarray}
h_s^i [ x^{-L_c-1}R_{n_0L_0}(x) Y^*_{\Omega_1 \omega_1}(\hat{\bf x}) ]&=&
[h_r^i(\Omega_1)x^{-L_c-1}R_{n_0L_0}(x)]Y^*_{\Omega_1 \omega_1}(\hat{\bf x})\,,
\label{eq127}
\end{eqnarray}
where on the right-hand-side, $h_r^i(\Omega_1)$ is understood to operate on all radial functions contained in the square brackets.
Furthermore, the product of $Y^*_{\ell m}(\hat{\bf x})$ and $Y^*_{\Omega_1 \omega_1}(\hat{\bf x})$ can be combined
into $Y_{\Omega_2 \omega_2}(\hat{\bf x})$. The application of (\ref{eq47}) again yields
\begin{eqnarray}
&&h_s^j [x^{-\ell-1}h_r^i(\Omega_1)x^{-L_c-1}R_{n_0L_0}(x)Y_{\Omega_2 \omega_2}(\hat{\bf x})  ]\nonumber\\
&&=
[h_r^j(\Omega_2)x^{-\ell-1}h_r^i(\Omega_1)x^{-L_c-1}R_{n_0L_0}(x)]Y_{\Omega_2 \omega_2}(\hat{\bf x})\,.
\label{eq128}
\end{eqnarray}
The last step is the integration over $d\Omega$ in (\ref{eq126}) for the product of three
remaining spherical harmonics $Y^*_{L_0M_0}(\hat{\bf x})$, $Y_{L_c'M_c'}(\hat{\bf x})$,
and $Y_{\Omega_2 \omega_2}(\hat{\bf x})$, which can be performed using (\ref{eq51}). We therefore arrive at
\begin{eqnarray}
&&w^{(3)}_{ij}(L_c',\ell,L_c) =\frac{2L_0+1}{(4\pi)^{3/2}}(L_c',\ell,L_c)^{1/2} \sum_{\Omega_1\Omega_2}
(\Omega_1,\Omega_2)
\left(
\begin{array}{ccc}
L_c & L_0 & \Omega_1 \\
0 & 0 & 0 \\
\end{array}
\right)
\left(
\begin{array}{ccc}
\ell & \Omega_1 & \Omega_2 \\
0 & 0 & 0 \\
\end{array}
\right)
\left(
\begin{array}{ccc}
L_c' & \Omega_2 & L_0 \\
0 & 0 & 0 \\
\end{array}
\right)\nonumber\\
&&\times G^{(3)}(\Omega_1,\Omega_2)\int_0^\infty dx x^{-L_c'+1}R_{n_0L_0}(x)
h_r^j(\Omega_2)\,[x^{-\ell-1}h_r^i(\Omega_1)\,x^{-L_c-1} R_{n_0L_0}(x)]\,,
\label{eq129}
\end{eqnarray}
where the angular coefficient $G^{(3)}$ is given by
\begin{eqnarray}
G^{(3)}(\Omega_1,\Omega_2) &\equiv& \sum_{M_c'mM_c}\sum_{\omega_1\omega_2}(-1)^{M_c'+M_c+M_0}
\left(
\begin{array}{ccc}
L_c' & \ell & L_c \\
-M_c' & m & M_c \\
\end{array}
\right)
\left(
\begin{array}{ccc}
L_c & L_0 & \Omega_1 \\
-M_c & M_0 & \omega_1 \\
\end{array}
\right)\nonumber\\
&\times&
\left(
\begin{array}{ccc}
\ell & \Omega_1 & \Omega_2 \\
m & \omega_1 & \omega_2 \\
\end{array}
\right)
\left(
\begin{array}{ccc}
L_c' & \Omega_2 & L_0 \\
M_c' & \omega_2 & -M_0 \\
\end{array}
\right)\,.
\label{eq130}
\end{eqnarray}
The above angular coefficient $G^{(3)}(\Omega_1,\Omega_2)$ can further be simplified using
the graphical method of angular momentum~\cite{zare}:
\begin{eqnarray}
G^{(3)}(\Omega_1,\Omega_2)&=& (-1)^{\ell+L_0}\frac{1}{2L_0+1}
\left\{
\begin{array}{ccc}
\Omega_2 & \Omega_1 & \ell \\
L_c & L_c' & L_0 \\
\end{array}
\right\}\,.
\label{eq131}
\end{eqnarray}
We finally obtain the following expression:
\begin{eqnarray}
&&w^{(3)}_{ij}(L_c',\ell,L_c) =\frac{(-1)^{\ell+L_0}}{(4\pi)^{3/2}}(L_c',\ell,L_c)^{1/2} \sum_{\Omega_1\Omega_2}
(\Omega_1,\Omega_2)
\left(
\begin{array}{ccc}
L_c & L_0 & \Omega_1 \\
0 & 0 & 0 \\
\end{array}
\right)
\left(
\begin{array}{ccc}
\ell & \Omega_1 & \Omega_2 \\
0 & 0 & 0 \\
\end{array}
\right)
\left(
\begin{array}{ccc}
L_c' & \Omega_2 & L_0 \\
0 & 0 & 0 \\
\end{array}
\right)\nonumber\\
&\times&
\left\{
\begin{array}{ccc}
\Omega_2 & \Omega_1 & \ell \\
L_c & L_c' & L_0 \\
\end{array}
\right\}
\int_0^\infty dx \,x^{-L_c'+1}R_{n_0L_0}(x)
\,h_r^j(\Omega_2)\,[x^{-\ell-1}\,h_r^i(\Omega_1)\,x^{-L_c-1} R_{n_0L_0}(x)]\,.
\label{eq132}
\end{eqnarray}
From the selection rule of the $3-j$ symbol, it is seen that
\begin{eqnarray}
L_c'+\ell + L_c = {\rm even}\,,
\label{eq133}
\end{eqnarray}
with the lowest value of 4. The correction $\Delta E_3$ may thus be expressed in the form
\begin{eqnarray}
\Delta E_3 &=& \sum_{i=0}^\infty \sum_{j=0}^\infty \sum_{s=4,6,8,\ldots}\delta e_3(i,j;s)\,,
\label{eq134}
\end{eqnarray}
where
\begin{eqnarray}
\delta e_3(i,j;s) &=&\sum_{\substack{L_c'\ell L_c\ge 1\\L_c'+\ell+L_c=s }}
(-1)^{i+j}
\frac{C_{L_c'}C_\ell C_{L_c}}{\sqrt{(L_c',L_c)}}w^{(3)}_c(i,j;L_c',\ell,L_c)w^{(3)}_{ij}(L_c',\ell,L_c)\,.
\label{eq135}
\end{eqnarray}
It is easy to show that
\begin{eqnarray}
\sum_s\delta e_3(i,j;s) &=& \sum_s\delta e_3(j,i;s)
\label{eq135a}
\end{eqnarray}
by noting that $w^{(3)}_c(i,j;L_c',\ell,L_c)=w^{(3)}_c(j,i;L_c,\ell,L_c')$ and all reduced matrix elements are real. Thus we can write the third-order energy correction as
\begin{eqnarray}
\Delta E_3 &=& \sum_{i=0}^\infty \sum_{s=4,6,8,\ldots}\delta e_3(i,i;s)
+2\sum_{i>j}^\infty \sum_{s=4,6,8,\ldots}\delta e_3(i,j;s)\,.
\label{eq135b}
\end{eqnarray}

We can similarly obtain $\delta e_3(i,j;s)$ for given $i$, $j$, and $s$, which are listed below:
\begin{eqnarray}
\delta e_3(0,0;4) &=&q_x^3\,a^7\,\pi^{3/2}
\bigg[\frac{16}{225}\sqrt{10}\,w^{(3)}_c(0,0;1,1,2)+\frac{8}{135}\sqrt{6}\,w^{(3)}_c(0,0;1,2,1)\bigg]
\langle x^{-7}\rangle_{n_0L_0}\,,
\label{eq136}
\end{eqnarray}
\begin{eqnarray}
\delta e_3(0,0;6) &=&-q_x^3\,a^9\,\pi^{3/2}
\bigg[\frac{16}{735}\sqrt{21}\,w^{(3)}_c(0,0;1,2,3)+\frac{16}{525}\sqrt{15}\,w^{(3)}_c(0,0;1,3,2)\nonumber\\
&+&\frac{16}{1225}\sqrt{35}\,w^{(3)}_c(0,0;2,1,3)+\frac{8}{875}\sqrt{14}\,w^{(3)}_c(0,0;2,2,2)
\bigg]
\langle x^{-9}\rangle_{n_0L_0}\,,
\label{eq137}
\end{eqnarray}
\begin{eqnarray}
\delta e_3(0,0;8) =&&q_x^3\,a^{11}\,\pi^{3/2}
\bigg[\frac{32}{567}\,w^{(3)}_c(0,0;1,3,4)+\frac{32}{1323}\sqrt{7}\,w^{(3)}_c(0,0;1,4,3)\nonumber\\
&+&\frac{16}{1575}\sqrt{14}\,w^{(3)}_c(0,0;2,2,4)+\frac{32}{3675}\sqrt{15}\,w^{(3)}_c(0,0;2,3,3)\nonumber\\
&+&\frac{8}{2625}\sqrt{70}\,w^{(3)}_c(0,0;2,4,2)+\frac{32}{3969}\sqrt{21}\,w^{(3)}_c(0,0;3,1,4)\nonumber\\
&+&\frac{16}{5145}\sqrt{21}\,w^{(3)}_c(0,0;3,2,3)
\bigg]
\langle x^{-11}\rangle_{n_0L_0}\,,
\label{eq138}
\end{eqnarray}
\begin{eqnarray}
\delta e_3(0,0;10) =&&-q_x^3\,a^{13}\,\pi^{3/2}
\bigg[\frac{16}{3267}\sqrt{55}\,w^{(3)}_c(0,0;1,4,5)+\frac{16}{891}\sqrt{5}\,w^{(3)}_c(0,0;1,5,4)\nonumber\\
&+&\frac{16}{12705}\sqrt{330}\,w^{(3)}_c(0,0;2,3,5)+\frac{32}{31185}\sqrt{385}\,w^{(3)}_c(0,0;2,4,4)\nonumber\\
&+&\frac{16}{8085}\sqrt{210}\,w^{(3)}_c(0,0;2,5,3)+\frac{16}{17787}\sqrt{462}\,w^{(3)}_c(0,0;3,2,5)\nonumber\\
&+&\frac{16}{4851}\sqrt{22}\,w^{(3)}_c(0,0;3,3,4)+\frac{8}{11319}\sqrt{154}\,w^{(3)}_c(0,0;3,4,3)\nonumber\\
&+&\frac{16}{9801}\sqrt{165}\,w^{(3)}_c(0,0;4,1,5)+\frac{16}{18711}\sqrt{77}\,w^{(3)}_c(0,0;4,2,4)
\bigg]
\langle x^{-13}\rangle_{n_0L_0}\,,
\label{eq139}
\end{eqnarray}
\begin{eqnarray}
\delta e_3(1,0;4) =&&-q_x^3\,a^{8}\,\pi^{3/2}
\bigg[\frac{8}{25}\sqrt{10}\,w^{(3)}_c(1,0;1,1,2)+\frac{16}{45}\sqrt{6}\,w^{(3)}_c(1,0;1,2,1)\nonumber\\
&+&\frac{16}{75}\sqrt{10}\,w^{(3)}_c(1,0;2,1,1)
\bigg]
\langle x^{-9}\rangle_{n_0L_0}\,,
\label{eq140}
\end{eqnarray}
\begin{eqnarray}
\delta e_3(1,0;6) =&&q_x^3\,a^{10}\,\pi^{3/2}
\bigg[\frac{128}{735}\sqrt{21}\,w^{(3)}_c(1,0;1,2,3)+\frac{32}{175}\sqrt{15}\,w^{(3)}_c(1,0;1,3,2)\nonumber\\
&+&  \frac{128}{1225}\sqrt{35}\,w^{(3)}_c(1,0;2,1,3)+\frac{96}{875}\sqrt{14}\,w^{(3)}_c(1,0;2,2,2)\nonumber\\
&+&    \frac{64}{525}\sqrt{15}\,w^{(3)}_c(1,0;2,3,1)+\frac{96}{1225}\sqrt{35}\,w^{(3)}_c(1,0;3,1,2)\nonumber\\
&+&    \frac{64}{735}\sqrt{21}\,w^{(3)}_c(1,0;3,2,1)\bigg]
\langle x^{-11}\rangle_{n_0L_0}\,,
\label{eq141}
\end{eqnarray}
\begin{eqnarray}
\delta e_3(1,0;8) =&&-q_x^3\,a^{12}\,\pi^{3/2}
\bigg[\frac{400}{567}\,w^{(3)}_c(1,0;1,3,4)+\frac{320}{1323}\sqrt{7}\,w^{(3)}_c(1,0;1,4,3)\nonumber\\
&+&  \frac{8}{63}\sqrt{14}\,w^{(3)}_c(1,0;2,2,4)+\frac{64}{735}\sqrt{15}\,w^{(3)}_c(1,0;2,3,3)\nonumber\\
&+&    \frac{8}{175}\sqrt{70}\,w^{(3)}_c(1,0;2,4,2)+\frac{400}{3969}\sqrt{21}\,w^{(3)}_c(1,0;3,1,4)\nonumber\\
&+&    \frac{64}{1029}\sqrt{21}\,w^{(3)}_c(1,0;3,2,3)+\frac{16}{245}\sqrt{15}\,w^{(3)}_c(1,0;3,3,2)\nonumber\\
&+&    \frac{160}{1323}\sqrt{7}\,w^{(3)}_c(1,0;3,4,1)+\frac{320}{3969}\sqrt{21}\,w^{(3)}_c(1,0;4,1,3)\nonumber\\
&+&    \frac{8}{105}\sqrt{14}\,w^{(3)}_c(1,0;4,2,2)+\frac{160}{567}\,w^{(3)}_c(1,0;4,3,1)
\bigg]
\langle x^{-13}\rangle_{n_0L_0}\,,
\label{eq142}
\end{eqnarray}
\begin{eqnarray}
&&\delta e_3(1,1;4) =-q_x^3\,a^{9}\,\pi^{3/2}\bigg\{Z_1
\bigg[\frac{4}{75}\sqrt{10}\,w^{(3)}_c(1,1;1,1,2)+\frac{4}{135}\sqrt{6}\,w^{(3)}_c(1,1;1,2,1)\bigg]\langle x^{-10}\rangle_{n_0L_0}\nonumber\\
&&-  \bigg[\frac{4}{25}\sqrt{10}\,(36+L_0(L_0+1))\,w^{(3)}_c(1,1;1,1,2)+\frac{16}{5}\sqrt{6}\,w^{(3)}_c(1,1;1,2,1)
\bigg]\langle x^{-11}\rangle_{n_0L_0}\bigg\}\,,
\label{eq143}
\end{eqnarray}
\begin{eqnarray}
\delta e_3(1,1;6) &=&-q_x^3\,a^{11}\,\pi^{3/2}\bigg\{-Z_1
\bigg[\frac{64}{3675}\sqrt{21}\,w^{(3)}_c(1,1;1,2,3)+\frac{16}{875}\sqrt{15}\,w^{(3)}_c(1,1;1,3,2)\nonumber\\
&&+\frac{96}{6125}\sqrt{35}\,w^{(3)}_c(1,1;2,1,3)+\frac{36}{4375}\sqrt{14}\,w^{(3)}_c(1,1;2,2,2)\bigg]\langle x^{-12}\rangle_{n_0L_0}\nonumber\\
&&+ \bigg[\frac{32}{3675}\sqrt{21}\,(440+7L_0(L_0+1))\,w^{(3)}_c(1,1;1,2,3)\nonumber\\
&&-\frac{32}{2625}\sqrt{15}\,(L_0(L_0+1)-330)\,w^{(3)}_c(1,1;1,3,2)\nonumber\\
&&+\frac{32}{6125}\sqrt{35}\,(660+13L_0(L_0+1))\,w^{(3)}_c(1,1;2,1,3)\nonumber\\
&&+\frac{48}{4375}\sqrt{14}\,(165+2L_0(L_0+1))\,w^{(3)}_c(1,1;2,2,2)
\bigg]\langle x^{-13}\rangle_{n_0L_0}\bigg\}\,,
\label{eq144}
\end{eqnarray}
\begin{eqnarray}
\delta e_3(2,0;4) &=&-q_x^3\,a^{9}\,\pi^{3/2}\bigg\{Z_1
\bigg[\frac{4}{75}\sqrt{10}\,w^{(3)}_c(2,0;1,1,2)+\frac{2}{27}\sqrt{6}\,w^{(3)}_c(2,0;1,2,1)\nonumber\\
&&+\frac{2}{45}\sqrt{10}\,w^{(3)}_c(2,0;2,1,1)\bigg]\langle x^{-10}\rangle_{n_0L_0}\nonumber\\
&&-\bigg[\frac{4}{25}\sqrt{10}\,(28+L_0(L_0+1))\,w^{(3)}_c(2,0;1,1,2)\nonumber\\
&&+\frac{2}{15}\sqrt{6}\,(28+L_0(L_0+1))\,w^{(3)}_c(2,0;1,2,1)\nonumber\\
&&+\frac{2}{25}\sqrt{10}\,(28+L_0(L_0+1))\,w^{(3)}_c(2,0;2,1,1)
\bigg]\langle x^{-11}\rangle_{n_0L_0}\bigg\}\,,
\label{eq145}
\end{eqnarray}
\begin{eqnarray}
\delta e_3(2,0;6) &=&-q_x^3\,a^{11}\,\pi^{3/2}\bigg\{-Z_1
\bigg[\frac{16}{735}\sqrt{21}\,w^{(3)}_c(2,0;1,2,3)+\frac{24}{875}\sqrt{15}\,w^{(3)}_c(2,0;1,3,2)\nonumber\\
&&+\frac{16}{1225}\sqrt{35}\,w^{(3)}_c(2,0;2,1,3)+\frac{72}{4375}\sqrt{14}\,w^{(3)}_c(2,0;2,2,2)\nonumber\\
&&+\frac{8}{375}\sqrt{15}\,w^{(3)}_c(2,0;2,3,1)+\frac{72}{6125}\sqrt{35}\,w^{(3)}_c(2,0;3,1,2)\nonumber\\
&&+\frac{8}{525}\sqrt{21}\,w^{(3)}_c(2,0;3,2,1)\bigg]\langle x^{-12}\rangle_{n_0L_0}\nonumber\\
&&+\bigg[\frac{64}{735}\sqrt{21}\,(45+L_0(L_0+1))\,w^{(3)}_c(2,0;1,2,3)\nonumber\\
&&+\frac{64}{875}\sqrt{15}\,(45+L_0(L_0+1))\,w^{(3)}_c(2,0;1,3,2)\nonumber\\
&&+\frac{64}{1225}\sqrt{35}\,(45+L_0(L_0+1))\,w^{(3)}_c(2,0;2,1,3)\nonumber\\
&&+\frac{192}{4375}\sqrt{14}\,(45+L_0(L_0+1))\,w^{(3)}_c(2,0;2,2,2)\nonumber\\
&&+\frac{32}{875}\sqrt{15}\,(45+L_0(L_0+1))\,w^{(3)}_c(2,0;2,3,1)\nonumber\\
&&+\frac{192}{6125}\sqrt{35}\,(45+L_0(L_0+1))\,w^{(3)}_c(2,0;3,1,2)\nonumber\\
&&+\frac{32}{1225}\sqrt{21}\,(45+L_0(L_0+1))\,w^{(3)}_c(2,0;3,2,1)
\bigg]\langle x^{-13}\rangle_{n_0L_0}\bigg\}\,,
\label{eq146}
\end{eqnarray}
\begin{eqnarray}
\delta e_3(3,0;4) &=&-q_x^3\,a^{10}\,\pi^{3/2}\bigg\{-Z_1
\bigg[\frac{56}{25}\sqrt{10}\,w^{(3)}_c(3,0;1,1,2)+\frac{184}{75}\sqrt{6}\,w^{(3)}_c(3,0;1,2,1)\nonumber\\
&&+\frac{184}{125}\sqrt{10}\,w^{(3)}_c(3,0;2,1,1)\bigg]\langle x^{-12}\rangle_{n_0L_0}\nonumber\\
&&+\bigg[\frac{64}{25}\sqrt{10}\,(35+3L_0(L_0+1))\,w^{(3)}_c(3,0;1,1,2)\nonumber\\
&&+\frac{128}{75}\sqrt{6}\,(35+3L_0(L_0+1))\,w^{(3)}_c(3,0;1,2,1)\nonumber\\
&&+\frac{128}{125}\sqrt{10}\,(35+3L_0(L_0+1))\,w^{(3)}_c(3,0;2,1,1)
\bigg]\langle x^{-13}\rangle_{n_0L_0}\bigg\}\,,
\label{eq147}
\end{eqnarray}
\begin{eqnarray}
\delta e_3(4,0;4) &&=-q_x^3\,a^{11}\,\pi^{3/2}\bigg\{-Z_1^2
\bigg[\frac{256}{825}\sqrt{10}\,w^{(3)}_c(4,0;1,1,2)+\frac{592}{1485}\sqrt{6}\,w^{(3)}_c(4,0;1,2,1)\nonumber\\
&&+\frac{592}{2475}\sqrt{10}\,w^{(3)}_c(4,0;2,1,1)\bigg]\langle x^{-13}\rangle_{n_0L_0}\nonumber\\
&&+Z_1\bigg[\frac{248}{275}\sqrt{10}\,(99+2L_0(L_0+1))\,w^{(3)}_c(4,0;1,1,2)\nonumber\\
&&+\frac{8}{165}\sqrt{6}\,(1661+29L_0(L_0+1))\,w^{(3)}_c(4,0;1,2,1)\nonumber\\
&&+\frac{8}{275}\sqrt{10}\,(1661+29L_0(L_0+1))\,w^{(3)}_c(4,0;2,1,1)\bigg]\langle x^{-14}\rangle_{n_0L_0}\nonumber\\
&&-\bigg[\frac{48}{25}\sqrt{10}\,(L_0^4+2L_0^3+179L_0^2+178L_0+1260)\,w^{(3)}_c(4,0;1,1,2)\nonumber\\
&&+\frac{16}{15}\sqrt{6}\,(L_0^4+2L_0^3+179L_0^2+178L_0+1260)\,w^{(3)}_c(4,0;1,2,1)\nonumber\\
&&+\frac{16}{25}\sqrt{10}\,(L_0^4+2L_0^3+179L_0^2+178L_0+1260)\,w^{(3)}_c(4,0;2,1,1)
\bigg]\langle x^{-15}\rangle_{n_0L_0}\bigg\}\,,
\label{eq148}
\end{eqnarray}
\begin{eqnarray}
\delta e_3(2,1;4) &&=-q_x^3\,a^{10}\,\pi^{3/2}\bigg\{-Z_1
\bigg[\frac{152}{125}\sqrt{10}\,w^{(3)}_c(2,1;1,1,2)+\frac{32}{25}\sqrt{6}\,w^{(3)}_c(2,1;1,2,1)\nonumber\\
&&+\frac{144}{125}\sqrt{10}\,w^{(3)}_c(2,1;2,1,1)\bigg]\langle x^{-12}\rangle_{n_0L_0}\nonumber\\
&&+\bigg[\frac{128}{125}\sqrt{10}\,(55+4L_0(L_0+1))\,w^{(3)}_c(2,1;1,1,2)\nonumber\\
&&+\frac{32}{75}\sqrt{6}\,(110+3L_0(L_0+1))\,w^{(3)}_c(2,1;1,2,1)\nonumber\\
&&+\frac{96}{125}\sqrt{10}\,(55+4L_0(L_0+1))\,w^{(3)}_c(2,1;2,1,1)\bigg]\langle x^{-13}\rangle_{n_0L_0}\bigg\}\,,
\label{eq149}
\end{eqnarray}
\begin{eqnarray}
\delta e_3(2,2;4) &&=-q_x^3\,a^{11}\,\pi^{3/2}\bigg\{-Z_1^2
\bigg[\frac{1216}{4125}\sqrt{10}\,w^{(3)}_c(2,2;1,1,2)+\frac{112}{675}\sqrt{6}\,w^{(3)}_c(2,2;1,2,1)
\bigg]\langle x^{-13}\rangle_{n_0L_0}\nonumber\\
&&+Z_1\bigg[\frac{16}{1375}\sqrt{10}\,(6919+128L_0(L_0+1))\,w^{(3)}_c(2,2;1,1,2)\nonumber\\
&&+\frac{32}{75}\sqrt{6}\,(88+L_0(L_0+1))\,w^{(3)}_c(2,2;1,2,1)
\bigg]\langle x^{-14}\rangle_{n_0L_0}\nonumber\\
&&-\bigg[\frac{176}{125}\sqrt{10}\,(L_0^4+2L_0^3+\frac{2069}{11}L_0^2+\frac{2058}{11}L_0+1560)\,w^{(3)}_c(2,2;1,1,2)\nonumber\\
&&+\frac{8}{25}\sqrt{6}\,(L_0^4+2L_0^3+129L_0^2+128L_0+2860)\,w^{(3)}_c(2,2;1,2,1)
\bigg]\langle x^{-15}\rangle_{n_0L_0}\bigg\}\,.
\label{eq150}
\end{eqnarray}

\subsubsection{Fourth-Order Energy}

The fourth-order energy correction can be evaluated according to
\begin{eqnarray}
\Delta E_4 &=& \Delta E^{(1)}_4+\Delta E^{(2)}_4\,,
\label{eq151}
\end{eqnarray}
where
\begin{eqnarray}
\label{eq152}
\Delta E^{(1)}_4 &\equiv& \langle \Psi_1 |v_{cx} |\Psi_2\rangle\,,\\
\label{eq153}
\Delta E^{(2)}_4 &\equiv& -\Delta E_2 \,\langle \Psi_1|\Psi_1\rangle\,.
\end{eqnarray}
In the above we have applied again the condition $\Delta E_1=0$;
also, $|\Psi_1\rangle$ is the first-order wave function correction given by
\begin{eqnarray}
|\Psi_1\rangle &=& \sum_{m}\frac{\langle \Psi_{m}|v_{cx}|\Psi_0\rangle}{(E_0-E_m)} |\Psi_{m}\rangle
\label{eq154}
\end{eqnarray}
and $|\Psi_2\rangle$ is the second-order
correction
\begin{eqnarray}
|\Psi_2\rangle &=& \sum_{k'n}\frac{\langle \Psi_{k'}|v_{cx}|\Psi_n\rangle
\langle \Psi_{n}|v_{cx}|\Psi_0\rangle}{(E_0-E_{k'})(E_0-E_n)} |\Psi_{k'}\rangle\,.
\label{eq155}
\end{eqnarray}
We focus on $\Delta E^{(1)}_4$ first. Substituting (\ref{eq154}) and (\ref{eq155}) into (\ref{eq152}) yields
\begin{eqnarray}
\Delta E^{(1)}_4 &=& \sum_{mnk'}\frac{\langle \Psi_{0}|v_{cx}|\Psi_m\rangle \langle \Psi_{m}|v_{cx}|\Psi_{k'}\rangle
\langle \Psi_{k'}|v_{cx}|\Psi_n\rangle \langle \Psi_{n}|v_{cx}|\Psi_0\rangle}
{(E_0-E_{m})(E_0-E_n)(E_0-E_{k'})}\,.
\label{eq156}
\end{eqnarray}
Let
\begin{eqnarray}
\label{eq157a}
\Psi_0 &=& \phi_0\chi_{n_0L_0M_0}\,,\\
\label{eq157b}
\Psi_m &=& \phi_{n_{c_1}L_{c_1}M_{c_1}} \chi_{n_{x_1}L_{x_1}M_{x_1}}\,,\\
\label{eq157c}
\Psi_{k'} &=& \phi_{n_{c_2}L_{c_2}M_{c_2}} \chi_{n_{x_2}L_{x_2}M_{x_2}}\,,\\
\label{eq157d}
\Psi_n &=& \phi_{n_{c_3}L_{c_3}M_{c_3}} \chi_{n_{x_3}L_{x_3}M_{x_3}}\,.
\end{eqnarray}
We first perform the following expansions according to (\ref{eq28})
\begin{eqnarray}
\frac{1}{E_0-E_m}&=&
\sum_{i=0}^\infty (-1)^{i+1}\frac{(\delta e_{n_{x_1}})^i}{(\delta \varepsilon_{n_{c_1}}(L_{c_1}))^{i+1}}\,
\label{eq158}
\end{eqnarray}
\begin{eqnarray}
\frac{1}{E_0-E_{k'}}&=&
\sum_{j=0}^\infty (-1)^{j+1}\frac{(\delta e_{n_{x_2}})^j}{(\delta \varepsilon_{n_{c_2}}(L_{c_2}))^{j+1}}\,
\label{eq159}
\end{eqnarray}
\begin{eqnarray}
\frac{1}{E_0-E_n}&=&
\sum_{k=0}^\infty (-1)^{k+1}\frac{(\delta e_{n_{x_3}})^k}{(\delta \varepsilon_{n_{c_3}}(L_{c_3}))^{k+1}}\,.
\label{eq160}
\end{eqnarray}
It should be pointed out that in making the above expansions, the necessary condition for these expansions to be valid
is that the excitation energies  $\delta \varepsilon_{n_{c_p}}(L_{c_p})\ne 0$ for $p=1,2,3$.
However, it is allowed for $\delta \varepsilon_{n_{c_2}}(L_{c_2})=0$, {\it i.e.},
$\phi_{n_{c_2}L_{c_2}M_{c_2}}=\phi_0$, because the intermediate state $\Psi_{k'}$ is connected
to another intermediate state $\Psi_{m}$ or $\Psi_{n}$ by $v_{cx}$. When this happens $E_0-E_{k'}=-\delta e_{n_{x_2}}$
and so a special treatment is needed for this case. We thus further split $\Delta E_4^{(1)}$ into two parts:
\begin{eqnarray}
\Delta E_4^{(1)} &=& \Delta E_{4a}^{(1)}+\Delta E_{4b}^{(1)}\,,
\label{eq161}
\end{eqnarray}
where the first term is for the case of $\delta \varepsilon_{n_{c_2}}(L_{c_2})\ne 0$, and the second term for
$\delta \varepsilon_{n_{c_2}}(L_{c_2})=0$. We deal with $\Delta E_{4a}^{(1)}$ first.

Substituting (\ref{eq157a})--(\ref{eq160}) into the right-hand-side of (\ref{eq156}) and evaluating
various matrix elements of $v_{cx}$ we obtain
\begin{eqnarray}
\Delta E_{4a}^{(1)} &=& \sum_{i,j,k=0}^\infty  (-1)^{i+j+k+1}
\sum_{L_{c_2}\ge 0}\sum_{L_{c_1},L_{c_3}\ge 1}\sum_{\ell_1,\ell_2\ge 1}
(-1)^{L_{c_2}}\frac{C_{L_{c_1}} C_{L_{c_3}} C_{\ell_1}C_{\ell_2}}{\sqrt{(L_{c_1},L_{{c_3}})}}\,\nonumber\\
&\times& w^{(4)}_c(i,j,k;L_{c_2},L_{c_1},\ell_1,\ell_2,L_{c_3})
w_{ijk}^{(4)}(L_{c_2},L_{c_1},\ell_1,\ell_2,L_{c_3})\,.
\label{eq162}
\end{eqnarray}
In the above, the quantity describing the core $w^{(4)}_c$ is defined by
{\small
\begin{eqnarray}
&&w^{(4)}_c(i,j,k;L_{c_2},L_{c_1},\ell_1,\ell_2,L_{c_3})\equiv\nonumber\\
&&\sum_{n_{c_1}n^*_{c_2}n_{c_3}}\frac
{ \langle \phi_0\|T_{L_{c_1}}\|\phi_{n_{c_1}L_{c_1}}\rangle
\langle \phi_{n_{c_1}L_{c_1}} \|T_{\ell_1} \| \phi_{n_{c_2}L_{c_2}}\rangle
\langle \phi_{n_{c_2}L_{c_2}} \|T_{\ell_2} \| \phi_{n_{c_3}L_{c_3}}\rangle
\langle \phi_{n_{c_3}L_{c_3}} \|T_{L_{c_3}} \| \phi_0\rangle}
{[\delta \varepsilon_{n_{c_1}}(L_{c_1})]^{i+1}
 [\delta \varepsilon_{n_{c_2}}(L_{c_2})]^{j+1}
 [\delta \varepsilon_{n_{c_3}}(L_{c_3})]^{k+1}}\,,
\label{eq163}
\end{eqnarray}
}
where $n^*_{c_2}$ indicates that the intermediate spectrum $\{\phi_{n_{c_2}L_{c_2}M_{c_2}}\}$ should exclude
the ground state of the core $\phi_0$.
It is easy to see that $w_c^{(4)}$ has the following symmetry:
\begin{eqnarray}
w^{(4)}_c(i,j,k;L_{c_2},L_{c_1},\ell_1,\ell_2,L_{c_3})
&=&w^{(4)}_c(k,j,i;L_{c_2},L_{c_3},\ell_2,\ell_1,L_{c_1})\,.
\label{eq164a}
\end{eqnarray}
The quantity describing the Rydberg electron $w_{ijk}^{(4)}$ is defined by
\begin{eqnarray}
&&w_{ijk}^{(4)}(L_{c_2},L_{c_1},\ell_1,\ell_2,L_{c_3})\equiv\nonumber\\
&&\sum_{M_{c_1}M_{c_2}M_{c_3}}\sum_{m_1m_2}(-1)^{M_{c_1}+M_{c_2}}
\left(
\begin{array}{ccc}
L_{c_1} & \ell_1 & L_{c_2} \\
-M_{c_1} & m_1 & M_{c_2} \\
\end{array}
\right)
\left(
\begin{array}{ccc}
L_{c_2} & \ell_2 & L_{c_3} \\
-M_{c_2} & m_2 & M_{c_3} \\
\end{array}
\right)\nonumber\\
&&\times \langle \chi_{n_0L_0M_0}|u_{L_{c_1}M_{c_1}} h_s^i u_{\ell_1 m_1}^* h_s^j u_{\ell_2m_2}^*
h_s^k u_{L_{c_3}M_{c_3}}^*|\chi_{n_0L_0M_0}\rangle\,.
\label{eq164}
\end{eqnarray}
One can further simplify $w_{ijk}^{(4)}$ by applying similar steps leading to (\ref{eq129}), arriving at
\begin{eqnarray}
&&w_{ijk}^{(4)}(L_{c_2},L_{c_1},\ell_1,\ell_2,L_{c_3}) =
\frac{2L_0+1}{(4\pi)^2}(L_{c_1},L_{c_3},\ell_1,\ell_2)^{1/2}
\sum_{\Omega_1\Omega_2\Omega_3}(\Omega_1,\Omega_2,\Omega_3)\nonumber\\
&&\times
\left(
\begin{array}{ccc}
L_{c_3} & L_0 & \Omega_1 \\
0 & 0 & 0 \\
\end{array}
\right)
\left(
\begin{array}{ccc}
\ell_2 & \Omega_1 & \Omega_2 \\
0 & 0 & 0 \\
\end{array}
\right)
\left(
\begin{array}{ccc}
\ell_1 & \Omega_2 & \Omega_3 \\
0 & 0 & 0 \\
\end{array}
\right)
\left(
\begin{array}{ccc}
L_0 & \Omega_3 & L_{c_1} \\
0 & 0 & 0 \\
\end{array}
\right)
G^{(4)}(\Omega_1,\Omega_2,\Omega_3)\nonumber\\
&&\times \int_0^\infty dx\,x^{-L_{c_1}+1}\,R_{n_0L_0}(x)\,
h_r^i(\Omega_3)\, x^{-\ell_1-1} \, h_r^j(\Omega_2) \, x^{-\ell_2-1}\, h_r^k(\Omega_1)\, x^{-L_{c_3}-1}\,R_{n_0L_0}(x)\,,
\label{eq165}
\end{eqnarray}
with $G^{(4)}$ being defined by
{\small
\begin{eqnarray}
&&G^{(4)}(\Omega_1,\Omega_2,\Omega_3) \equiv\sum_{M_{c_1}M_{c_2}M_{c_3}}\sum_{m_1m_2}\sum_{\omega_1\omega_2\omega_3}
(-1)^{M_{c_2}+M_{c_3}+m_1}
\left(
\begin{array}{ccc}
L_{c_1} & \ell_1 & L_{c_2} \\
-M_{c_1} & m_1 & M_{c_2} \\
\end{array}
\right)
\left(
\begin{array}{ccc}
L_{c_2} & \ell_2 & L_{c_3} \\
-M_{c_2} & m_2 & M_{c_3} \\
\end{array}
\right)\nonumber\\
&&\times
\left(
\begin{array}{ccc}
L_{c_3} & L_0 & \Omega_1 \\
-M_{c_3} & M_0 & \omega_1 \\
\end{array}
\right)
\left(
\begin{array}{ccc}
\ell_2 & \Omega_1 & \Omega_2 \\
m_2 & \omega_1 & \omega_2 \\
\end{array}
\right)
\left(
\begin{array}{ccc}
\ell_1 & \Omega_2 & \Omega_3 \\
-m_1 & \omega_2 & \omega_3 \\
\end{array}
\right)
\left(
\begin{array}{ccc}
L_0 & \Omega_3 & L_{c_1} \\
M_0 & \omega_3 & -M_{c_1} \\
\end{array}
\right)\,.
\label{eq166}
\end{eqnarray}
}
The use of the graphical method of angular momentum leads to~\cite{zare}
\begin{eqnarray}
G^{(4)}(\Omega_1,\Omega_2,\Omega_3) &=&
(-1)^{\ell_1+\ell_2}\frac{1}{2L_0+1}
\left\{
\begin{array}{ccc}
\Omega_3 & \Omega_2 & \ell_1 \\
L_{c_2} & L_{c_1} & L_0 \\
\end{array}
\right\}
\left\{
\begin{array}{ccc}
\Omega_2 & \Omega_1 & \ell_2 \\
L_{c_3} & L_{c_2} & L_0 \\
\end{array}
\right\}
\,.
\label{eq167}
\end{eqnarray}
We finally have
\begin{eqnarray}
&&w_{ijk}^{(4)}(L_{c_2},L_{c_1},\ell_1,\ell_2,L_{c_3}) =
\frac{(-1)^{\ell_1+\ell_2}}{(4\pi)^2}(L_{c_1},L_{c_3},\ell_1,\ell_2)^{1/2}
\sum_{\Omega_1\Omega_2\Omega_3}(\Omega_1,\Omega_2,\Omega_3)
\left(
\begin{array}{ccc}
L_{c_3} & L_0 & \Omega_1 \\
0 & 0 & 0 \\
\end{array}
\right)
\nonumber\\
&&\times\left(
\begin{array}{ccc}
\ell_2 & \Omega_1 & \Omega_2 \\
0 & 0 & 0 \\
\end{array}
\right)
\left(
\begin{array}{ccc}
\ell_1 & \Omega_2 & \Omega_3 \\
0 & 0 & 0 \\
\end{array}
\right)
\left(
\begin{array}{ccc}
L_0 & \Omega_3 & L_{c_1} \\
0 & 0 & 0 \\
\end{array}
\right)
\left\{
\begin{array}{ccc}
\Omega_3 & \Omega_2 & \ell_1 \\
L_{c_2} & L_{c_1} & L_0 \\
\end{array}
\right\}
\left\{
\begin{array}{ccc}
\Omega_2 & \Omega_1 & \ell_2 \\
L_{c_3} & L_{c_2} & L_0 \\
\end{array}
\right\}
\nonumber\\
&&\times \int_0^\infty dx\,x^{-L_{c_1}+1}\,R_{n_0L_0}(x)\,
h_r^i(\Omega_3)\, x^{-\ell_1-1} \, h_r^j(\Omega_2) \, x^{-\ell_2-1}\, h_r^k(\Omega_1)\, x^{-L_{c_3}-1}\,R_{n_0L_0}(x)\,.
\label{eq168}
\end{eqnarray}
From the four $3-j$ symbols in (\ref{eq168}) one can see that $L_{c_1}+\ell_1+\ell_2+L_{c_3}$
must be even with the lowest value of 4. We thus rewrite $\Delta E_{4a}^{(1)}$ in the form
\begin{eqnarray}
\Delta E_{4a}^{(1)} &=& \sum_{i,j,k=0}^\infty  \sum_{s=4,6,8,\ldots}\delta e_4(i,j,k;s)\,,
\label{eq169}
\end{eqnarray}
where
\begin{eqnarray}
\delta e_4(i,j,k;s) &=& \sum_{L_{c_2}\ge 0}\sum_{\substack{L_{c_1},\ell_1,\ell_2,L_{c_3}\ge 1\\L_{c_1}+\ell_1+\ell_2+L_{c_3}=s }}
(-1)^{i+j+k+1+L_{c_2}}
\frac{C_{L_{c_1}} C_{L_{c_3}} C_{\ell_1}C_{\ell_2}}{\sqrt{(L_{c_1},L_{{c_3}})}}\,\nonumber \\
&\times& w^{(4)}_c(i,j,k;L_{c_2},L_{c_1},\ell_1,\ell_2,L_{c_3})
w_{ijk}^{(4)}(L_{c_2},L_{c_1},\ell_1,\ell_2,L_{c_3})\,.
\label{eq170}
\end{eqnarray}

We now list $\delta e_4(i,j,k;s)$ below up to the order of $\langle x^{-10}\rangle_{n_0L_0}$, where the symmetry condition (\ref{eq164a}) is applied:
\begin{eqnarray}
\delta e_4(0,0,0;4) =&-&q_x^4\,a^8\,\pi^2\bigg[\frac{16}{81}w_c^{(4)}(0,0,0;0,1,1,1,1)\nonumber\\
&+&\frac{32}{405}w_c^{(4)}(0,0,0;2,1,1,1,1)\bigg]\langle x^{-8}\rangle_{n_0L_0}\,,
\label{eq171}
\end{eqnarray}
\begin{eqnarray}
\delta e_4(1,0,0;4) =&&q_x^4\,a^9\,\pi^2\bigg[\frac{112}{81}w_c^{(4)}(0,0,1;0,1,1,1,1)\nonumber\\
&+&\frac{224}{405}w_c^{(4)}(0,0,1;2,1,1,1,1)\bigg]\langle x^{-10}\rangle_{n_0L_0}\,,
\label{eq172}
\end{eqnarray}
\begin{eqnarray}
\delta e_4(0,1,0;4) =&&q_x^4\,a^9\,\pi^2\bigg[\frac{128}{81}w_c^{(4)}(0,1,0;0,1,1,1,1)\nonumber\\
&+&\frac{352}{405}w_c^{(4)}(0,1,0;2,1,1,1,1)\bigg]\langle x^{-10}\rangle_{n_0L_0}\,,
\label{eq173}
\end{eqnarray}
\begin{eqnarray}
\delta e_4(0,0,1;4) =&&q_x^4\,a^9\,\pi^2\bigg[\frac{112}{81}w_c^{(4)}(0,0,1;0,1,1,1,1)\nonumber\\
&+&\frac{224}{405}w_c^{(4)}(0,0,1;2,1,1,1,1)\bigg]\langle x^{-10}\rangle_{n_0L_0}\,,
\label{eq174}
\end{eqnarray}
\begin{eqnarray}
\delta e_4(0,0,0;6) =&&q_x^4\,a^{10}\,\pi^2\bigg[\frac{32}{945}\sqrt{6}\,w_c^{(4)}(0,0,0;2,1,1,3,1)\nonumber\\
&+& \frac{32}{675}w_c^{(4)}(0,0,0;1,1,2,2,1)\nonumber\\
&+& \frac{16}{525}w_c^{(4)}(0,0,0;3,1,2,2,1)\nonumber\\
&+& \frac{32}{225}w_c^{(4)}(0,0,0;0,1,1,2,2)\nonumber\\
&+& \frac{64}{7875}\sqrt{35}\,w_c^{(4)}(0,0,0;2,1,1,2,2)\nonumber\\
&+& \frac{64}{3375}\sqrt{15}\,w_c^{(4)}(0,0,0;1,2,1,2,1)\nonumber\\
&+& \frac{32}{2625}\sqrt{15}\,w_c^{(4)}(0,0,0;3,2,1,2,1)\nonumber\\
&+& \frac{32}{2205}\sqrt{14}\,w_c^{(4)}(0,0,0;2,1,1,1,3)\nonumber\\
&+& \frac{32}{1125}w_c^{(4)}(0,0,0;1,2,1,1,2)\nonumber\\
&+& \frac{16}{875}w_c^{(4)}(0,0,0;3,2,1,1,2)\bigg]\langle x^{-10}\rangle_{n_0L_0}\,.
\label{eq175}
\end{eqnarray}

Let us consider $\langle\Psi_1|\Psi_1\rangle$ in (\ref{eq153}), which can be expressed in the form
\begin{eqnarray}
\langle\Psi_1|\Psi_1\rangle &=& \sum_n \frac{\langle\Psi_0|v_{cx}|\Psi_n\rangle\langle\Psi_n|v_{cx}|\Psi_0\rangle}{(E_0-E_n)^2}\,.
\label{eq178}
\end{eqnarray}
Assuming
\begin{eqnarray}
\label{eq178a}
\Psi_0 &=& \phi_0\chi_{n_0L_0M_0}\,,\\
\label{eq178b}
\Psi_n &=& \phi_{n_{c}L_{c}M_{c}} \chi_{n_{x}L_{x}M_{x}}\,,
\end{eqnarray}
and using the expansion of $1/(E_0-E_n)$ in (\ref{eq28}), $\langle\Psi_1|\Psi_1\rangle$ can be reduced to
\begin{eqnarray}
\langle\Psi_1|\Psi_1\rangle &=& \sum_{i,j=0}^\infty (-1)^{i+j}\sum_{L_c\ge 1}\frac{C_{L_c}^2(2L_c+1)2^{i+j-2}}{\pi}
\alpha(i+j+1,L_c)w_{i+j}^{(2)}(L_c)\,,
\label{eq179}
\end{eqnarray}
where $\alpha(i+j+1,L_c)$ and $w_{i+j}^{(2)}(L_c)$ are defined in (\ref{eq40}) and (\ref{eq37}) respectively.
Since $\langle\Psi_1|\Psi_1\rangle$ depends on $i$ and $j$ through $i+j$, we can apply the following transformation
\begin{eqnarray}
\sum_{j=0}^\infty \sum_{i=0}^\infty f(i+j)=\sum_{k=0}^\infty\sum_{i=0}^k f(k)=\sum_{k=0}^\infty (k+1) f(k)\,,
\label{eq180}
\end{eqnarray}
to (\ref{eq179}) resulting in
\begin{eqnarray}
\langle\Psi_1|\Psi_1\rangle &=&q_x^2\,\sum_{i=0}^\infty(-1)^i(i+1)\sum_{L_c\ge 1}\frac{2^{i+2}\pi a^{2L_c+2}}{2L_c+1}
\alpha(i+1,L_c)w_i^{(2)}(L_c)\,.
\label{eq181}
\end{eqnarray}
Comparing to (\ref{eq42}) one can see that $\langle\Psi_1|\Psi_1\rangle$ has the same expression as $\Delta E_2$,
provided that $-2(i+1)\alpha(i+1,L_c)$ is replaced by $\alpha(i,L_c)$. Thus if we set
\begin{eqnarray}
\langle\Psi_1|\Psi_1\rangle &=& \sum_{i=0}^\infty \sum_{L_c \ge 1} e_p(i,L_c)\,,
\label{eq182}
\end{eqnarray}
according to (\ref{eq45x}), (\ref{eq53x}), and (\ref{eq101x}), we have the following specific expressions:
\begin{eqnarray}
e_p(0,L_c) &=& q_x^2\,a^{2L_c+2}\,\beta_{L_c}\,\langle x^{-2L_c-2}\rangle_{n_0L_0}\,,
\label{eq183}
\end{eqnarray}
\begin{eqnarray}
e_p(1,L_c) &=& -2\,q_x^2\,a^{2L_c+3}(L_c+1)(2L_c+1)\,\gamma_{L_c}\,\langle x^{-2L_c-4}\rangle_{n_0L_0}\,,
\label{eq184}
\end{eqnarray}
\begin{eqnarray}
&&e_p(2,L_c) = -6\,q_x^2\,a^{2L_c+4}\,\delta_{L_c}(L_c+1)^2\bigg\{\frac{2Z_1}{2L_c+3}\langle x^{-2L_c-5}\rangle_{n_0L_0}\nonumber\\
&&
-(L_c+2)(2L_c+1)\bigg(1+\frac{L_0(L_0+1)}{(L_c+1)(2L_c+3)}\bigg)\langle x^{-2L_c-6}\rangle_{n_0L_0}\bigg\}\,.
\label{eq185}
\end{eqnarray}
Using these results, one obtains the following correction of (\ref{eq153}) up to order $O(x^{-10})$
\begin{eqnarray}
\Delta E_4^{(2)} &=& \frac{1}{2}\,q_x^4\,a^8\,\alpha_1\,\beta_1\,\langle x^{-4}\rangle_{n_0L_0}^2
+\frac{1}{2}\,q_x^4\,\,\bigg[a^{10}\,\alpha_1\,\beta_2+a^{10}\,\beta_1\,\alpha_2\nonumber\\
&-&12\,a^9\,\alpha_1\,\gamma_1-6\,a^9\,\beta_1^2\bigg]
\,\langle x^{-4}\rangle_{n_0L_0}\,\langle x^{-6}\rangle_{n_0L_0}+O(x^{-11})\,.
\label{eq186}
\end{eqnarray}

Finally we consider $\Delta E_{4b}^{(1)}$ in (\ref{eq161}), which corresponds to the case of
$\phi_{n_{c_2}L_{c_2}M_{c_2}}=\phi_0$ and thus $E_0-E_{k'}=-\delta e_{n_{x_2}}=-(e_{n_{x_2}}-e_{n_0})$ in (\ref{eq156}). After
evaluating relevant matrix elements of $v_{cx}$, we obtain the following expression
\begin{eqnarray}
\Delta E_{4b}^{(1)} &=& \sum_{i,k=0}^\infty (-1)^{i+k+1}\frac{2^{i+k-6}}{\pi^2}\sum_{L_{c_1},L_{c_3}\ge 1}C^2_{L_{c_1}} C^2_{L_{c_3}}(L_{c_1},L_{c_3})W_g(i,k;L_{c_1},L_{c_3})\nonumber\\
&\times&\alpha(i,L_{c_1})\alpha(k,L_{c_3}),
\label{eq187}
\end{eqnarray}
where $\alpha(i,L_c)$ is the $2^{L_c}$-pole ``generalized polarizability" defined in (\ref{eq40}) and
\begin{eqnarray}
W_g(i,k;L_{c_1},L_{c_3}) = \sum_{M_{c_1}M_{c_3}}\langle\chi_{n_0L_0M_0}|u_{L_{c_1}M_{c_1}} h_s^i u^*_{L_{c_1}M_{c_1}}
\hat{G}(n_0) u_{L_{c_3}M_{c_3}} h_s^k u^*_{L_{c_3}M_{c_3}}|\chi_{n_0L_0M_0}\rangle &&
\label{eq188}
\end{eqnarray}
with $\hat{G}(n_0)$ being the reduced Schr\"{o}dinger-Coulomb Green function defined in~\cite{Swainson,Swainson2}
\begin{eqnarray}
\hat{G}(n_0) &\equiv& \sum_{n_{x_2}L_{x_2}M_{x_2}}
\frac{|\chi_{ n_{x_2}L_{x_2}M_{x_2}}\rangle\langle \chi_{n_{x_2}L_{x_2}M_{x_2}}|}
{e_{n_{x_2}}-e_{n_0}}\,.
\label{eq189}
\end{eqnarray}
It should be mentioned that in (\ref{eq189}), the sum is over all states, including the continuum, with $e_{n_{x_2}}\ne e_{n_0}$.
By taking the complex conjugate of $W_g(i,k;L_{c_1},L_{c_3})$ and noting that it is real, one arrives at the following relation
\begin{eqnarray}
W_g(i,k;L_{c_1},L_{c_3})&=&W_g(k,i;L_{c_3},L_{c_1})\,.
\label{eq190}
\end{eqnarray}
Using (\ref{eq44x}) one can see that
\begin{eqnarray}
W_g(0,0;L_{c_1},L_{c_3})&=& \frac{(L_{c_1},L_{c_3})}{16\pi^2}S_{2L_{c_1},2L_{c_3}}(n_0,L_0)\,,
\label{eq191}
\end{eqnarray}
where
\begin{eqnarray}
S_{i,j}(n_0,L_0)&=&\langle \chi_{n_0 L_0 M_0} | x^{-i-2}\hat{G}(e_{n_0}) x^{-j-2}|\chi_{n_0 L_0 M_0}\rangle\,.
\label{eq192}
\end{eqnarray}
In general, $W_g$ can further be recast into
\begin{eqnarray}
W_g(i,k;L_{c_1},L_{c_3}) &=& \langle\chi_{n_0L_0M_0}|\mathcal{\hat{U}}_i(L_{c_1})\hat{G}(e_{n_0})\mathcal{\hat{U}}_k(L_{c_3})
|\chi_{n_0L_0M_0}\rangle\nonumber\\
&=& \langle\chi_{n_0L_0M_0}|\mathcal{\hat{U}}_i(L_{c_1})|g_k(L_{c_3})\rangle\,,
\label{eq194}
\end{eqnarray}
where
\begin{eqnarray}
|g_k(L_{c_3})\rangle&=& \hat{G}(e_{n_0})\mathcal{\hat{U}}_k(L_{c_3})|\chi_{n_0L_0M_0}\rangle\nonumber\\
&=& \sum_{n_{x2}L_{x_2}M_{x_2}}\frac{|\chi_{n_{x2}L_{x_2}M_{x_2}}\rangle\langle\chi_{n_{x2}L_{x_2}M_{x_2}}
|\mathcal{\hat{U}}_k(L_{c_3})|\chi_{n_{0}L_{0}M_{0}}\rangle}
{e_{n_{x_2}}-e_{n_0}}
\label{eq195}
\end{eqnarray}
and $\mathcal{\hat{U}}_i(\ell)$ is defined in (\ref{eq36a}).
The above defined $|g_k(L_{c_3})\rangle$ may be interpreted as
the first-order wave function ``correction" due to the ``perturbation" $-\mathcal{\hat{U}}_k(L_{c_3})$, thus satisfying the following equation
\begin{eqnarray}
h_s|g_k(L_{c_3})\rangle&=& \mathcal{\hat{U}}_k(L_{c_3}) |\chi_{n_{0}L_{0}M_{0}}\rangle-
\langle \chi_{n_{0}L_{0}M_{0}} |\mathcal{\hat{U}}_k(L_{c_3})|\chi_{n_{0}L_{0}M_{0}}\rangle |\chi_{n_{0}L_{0}M_{0}}\rangle\,.
\label{eq196}
\end{eqnarray}
This equation can be considered as the reduction formula for $h_s$ acting on $|g_k(L_{c_3})\rangle$, where the right-hand-side
of (\ref{eq196}) does not involve the Green function.

Next consider the following case
\begin{eqnarray}
W_g(1,k;L_{c_1},L_{c_3})&=& \sum_{M_{c_1}}\langle \chi_{n_{0}L_{0}M_{0}} | u_{L_{c_1}M_{c_1}} h_s u^*_{L_{c_1}M_{c_1}}
|g_k(L_{c_3})\rangle\,.
\label{eq197}
\end{eqnarray}
In order to simply the above expression, we try to move $h_s$ to act on $|g_k(L_{c_3})\rangle$ directly so that
(\ref{eq196}) can be applied. Since
\begin{eqnarray}
u_{L_{c_1}M_{c_1}} h_s u^*_{L_{c_1}M_{c_1}}&=& u_{L_{c_1}M_{c_1}} [h_s,u^*_{L_{c_1}M_{c_1}}]
+u_{L_{c_1}M_{c_1}} u^*_{L_{c_1}M_{c_1}} h_s\nonumber\\
&=& -a \, u_{L_{c_1}M_{c_1}} \nabla u^*_{L_{c_1}M_{c_1}} \cdot \nabla + u_{L_{c_1}M_{c_1}} u^*_{L_{c_1}M_{c_1}} h_s
\label{eq198}
\end{eqnarray}
according to (\ref{eq50a}), we have
\begin{eqnarray}
W_g(1,k;L_{c_1},L_{c_3}) &=& a\, \sum_{M_{c_1}}\int d^3 x (\nabla \chi^*_{n_{0}L_{0}M_{0}})\cdot (\nabla u^*_{L_{c_1}M_{c_1}}) u_{L_{c_1}M_{c_1}}\, g_k(L_{c_3})\nonumber\\
&+&\frac{a}{2} \langle\chi_{n_{0}L_{0}M_{0}}|(\nabla^2 \sum_{M_{c_1}}u_{L_{c_1}M_{c_1}}u^*_{L_{c_1}M_{c_1}})|g_k(L_{c_3})\rangle\nonumber\\
&+&\langle \chi_{n_{0}L_{0}M_{0}}|\sum_{M_{c_1}} u_{L_{c_1}M_{c_1}} u^*_{L_{c_1}M_{c_1}} h_s|g_k(L_{c_3})\rangle\,.
\label{eq199}
\end{eqnarray}
In the above, we have performed an integration by parts and applied $\nabla^2 u_{\ell m}=0$ and
2$\nabla u_{\ell m}\cdot \nabla u^*_{\ell m}=\nabla^2 (u_{\ell m}u^*_{\ell m}) $. On the other hand, using
\begin{eqnarray}
u_{L_{c_1}M_{c_1}} h_s u^*_{L_{c_1}M_{c_1}}&=&  [u_{L_{c_1}M_{c_1}},h_s]u^*_{L_{c_1}M_{c_1}}
+h_s u_{L_{c_1}M_{c_1}}u^*_{L_{c_1}M_{c_1}}\nonumber\\
&=& a\,(\nabla u_{L_{c_1}M_{c_1}})\cdot \nabla u^*_{L_{c_1}M_{c_1}}+h_s u_{L_{c_1}M_{c_1}}u^*_{L_{c_1}M_{c_1}}
\label{eq200}
\end{eqnarray}
and noting that $\langle \chi_{n_{0}L_{0}M_{0}}|h_s=0 $, we have
\begin{eqnarray}
W_g(1,k;L_{c_1},L_{c_3}) &=& -a\,\sum_{M_{c_1}}\int d^3x (\nabla \chi^*_{n_{0}L_{0}M_{0}} )\cdot
(\nabla u_{L_{c_1}M_{c_1}} ) u^*_{L_{c_1}M_{c_1}} \,g_k(L_{c_3})
\label{eq201}
\end{eqnarray}
after performing an integration by parts. Furthermore, it is easy to verify that
\begin{eqnarray}
\sum_m (\nabla u_{\ell m}) u^*_{\ell m}&=& \sum_m u_{\ell m} (\nabla u^*_{\ell m})
\label{eq202}
\end{eqnarray}
according to (\ref{eq51x}). Therefore, by adding (\ref{eq199}) and (\ref{eq201})
and using the formula $\sum_m u_{\ell m}u^*_{\ell m}=x^{-2\ell-2}\,(2\ell+1)/(4\pi)$ we arrive at
\begin{eqnarray}
W_g(1,k;L_{c_1},L_{c_3})&=& \frac{2L_{c_1}+1}{8\pi}
\bigg[\frac{a}{2}\,\langle \chi_{n_{0}L_{0}M_{0}}|\nabla^2 x^{-2L_{c_1}-2}|g_k(L_{c_3})\rangle\nonumber\\
&+&\langle \chi_{n_{0}L_{0}M_{0}}|x^{-2L_{c_1}-2}\,h_s|g_k(L_{c_3})\rangle
\bigg]\nonumber\\
&=& \frac{2L_{c_1}+1}{8\pi}\bigg[a\,(L_{c_1}+1)(2L_{c_1}+1)\langle \chi_{n_{0}L_{0}M_{0}}|x^{-2L_{c_1}-4}|g_k(L_{c_3})\rangle\nonumber\\
&+& \langle \chi_{n_{0}L_{0}M_{0}}|x^{-2L_{c_1}-2}\,\mathcal{\hat{U}}_k(L_{c_3})|\chi_{n_{0}L_{0}M_{0}}\rangle\nonumber\\
&-& \langle x^{-2L_{c_1}-2}\rangle_{n_0L_0}
\langle \chi_{n_{0}L_{0}M_{0}}|\mathcal{\hat{U}}_k(L_{c_3})|\chi_{n_{0}L_{0}M_{0}}\rangle
\bigg]\,,
\label{eq203}
\end{eqnarray}
where (\ref{eq196}) has been used. Consider the case of $k=0$. Since
\begin{eqnarray}
\mathcal{\hat{U}}_0(L_{c_3})&=&\frac{2L_{c_3}+1}{4\pi} x^{-2L_{c_3}-2}\,,\\
|g_0(L_{c_3})\rangle&=&\frac{2L_{c_3}+1}{4\pi} \hat{G}(e_{n_0}) x^{-2L_{c_3}-2} |\chi_{n_{0}L_{0}M_{0}}\rangle\,,
\label{eq204}
\end{eqnarray}
we have the following expression
\begin{eqnarray}
W_g(1,0;L_{c_1},L_{c_3})&=&\frac{(L_{c_1},L_{c_3})}{32\pi^2}\bigg[
a(L_{c_1}+1)(2L_{c_1}+1)S_{2L_{c_1}+2,2L_{c_3}}(n_0,L_0)\nonumber\\
&+&\langle x^{-2(L_{c_1}+L_{c_3})-4}\rangle_{n_0L_0}
-\langle x^{-2L_{c_1}-2}\rangle_{n_0L_0}\langle x^{-2L_{c_3}-2}\rangle_{n_0L_0}
\bigg]\,.
\label{eq205}
\end{eqnarray}
It is noted that $W_g(1,k;L_{c_1},L_{c_3})$ in (\ref{eq203}) can further be expressed according to
\begin{eqnarray}
W_g(1,k;L_{c_1},L_{c_3})&=&\frac{a(L_{c_1}+1)(2L_{c_1}+1)^2}{2(2L_{c_1}+3)}W_g(0,k;L_{c_1}+1,L_{c_3})\nonumber\\
&+&\frac{1}{2}w^{(4)}_{0k}(L_{c_1},L_{c_3})-\frac{2L_{c_1}+1}{8\pi}w_k^{(2)}(L_{c_3})\,\langle x^{-2L_{c_1}-2}\rangle_{n_0L_0}\,,
\label{eq206}
\end{eqnarray}
where
\begin{eqnarray}
w^{(4)}_{ij}(\ell,\ell')&\equiv& \langle \chi_{n_{0}L_{0}M_{0}} |\mathcal{\hat{U}}_i(\ell)\mathcal{\hat{U}}_j(\ell')|\chi_{n_{0}L_{0}M_{0}}\rangle\,,
\label{eq207}
\end{eqnarray}
and $w_i^{(2)}(\ell)=\langle \chi_{n_{0}L_{0}M_{0}} |\mathcal{\hat{U}}_i(\ell)|\chi_{n_{0}L_{0}M_{0}}\rangle$ is defined in
(\ref{eq37}). Since $w^{(4)}_{ij}$ is real, it is seen that $w^{(4)}_{ij}(\ell,\ell')=w^{(4)}_{ji}(\ell',\ell)$. Applying a
similar procedure leading to (\ref{eq132}), we arrive at
\begin{eqnarray}
w^{(4)}_{ij}(\ell,\ell')&=&
\frac{(\ell,\ell')}{16\pi^2} \sum_{\Omega_1\Omega_2}
(\Omega_1,\Omega_2)
\left(
\begin{array}{ccc}
\ell' & L_0 & \Omega_1 \\
0 & 0 & 0 \\
\end{array}
\right)^2
\left(
\begin{array}{ccc}
\ell & L_0 & \Omega_2 \\
0 & 0 & 0 \\
\end{array}
\right)^2\nonumber\\
&\times&\int_0^\infty dx \,x^{-\ell+1}R_{n_0L_0}(x)
\,h_r^i(\Omega_2)\,[x^{-\ell-\ell'-2}\,h_r^j(\Omega_1)\,x^{-\ell'-1} R_{n_0L_0}(x)]\,.
\label{eq208}
\end{eqnarray}
We list some special values for $w^{(4)}_{ij}(\ell,\ell')$ below:
\begin{eqnarray}
w^{(4)}_{00}(\ell,\ell')&=& \frac{(\ell,\ell')}{16\pi^2}\langle x^{-2(\ell+\ell')-4}\rangle_{n_0L_0}\,,
\label{eq209}
\end{eqnarray}
\begin{eqnarray}
w^{(4)}_{01}(1,1)&=& \frac{63a}{16\pi^2}\langle x^{-10}\rangle_{n_0L_0}\,,
\label{eq210}
\end{eqnarray}
\begin{eqnarray}
w^{(4)}_{01}(2,1)&=& \frac{135a}{16\pi^2}\langle x^{-12}\rangle_{n_0L_0}\,,
\label{eq211}
\end{eqnarray}
\begin{eqnarray}
w^{(4)}_{01}(1,2)&=& \frac{405a}{32\pi^2}\langle x^{-12}\rangle_{n_0L_0}\,,
\label{eq212}
\end{eqnarray}
\begin{eqnarray}
w^{(4)}_{01}(2,2)&=& \frac{825a}{32\pi^2}\langle x^{-14}\rangle_{n_0L_0}\,,
\label{eq213}
\end{eqnarray}
\begin{eqnarray}
w^{(4)}_{02}(1,1)&=& -\frac{3Z_1a^2}{4\pi^2}\langle x^{-11}\rangle_{n_0L_0}
+\frac{21a^2(L_0^2+L_0+36)}{16\pi^2}\langle x^{-12}\rangle_{n_0L_0}\,,
\label{eq213a}
\end{eqnarray}
\begin{eqnarray}
w^{(4)}_{02}(1,2)&=& -\frac{315Z_1a^2}{176\pi^2}\langle x^{-13}\rangle_{n_0L_0}
+\frac{405a^2(L_0^2+L_0+55)}{88\pi^2}\langle x^{-14}\rangle_{n_0L_0}\,,
\label{eq213b}
\end{eqnarray}
\begin{eqnarray}
w^{(4)}_{02}(2,1)&=& -\frac{15Z_1a^2}{11\pi^2}\langle x^{-13}\rangle_{n_0L_0}
+\frac{405a^2(L_0^2+L_0+55)}{176\pi^2}\langle x^{-14}\rangle_{n_0L_0}\,,
\label{eq213c}
\end{eqnarray}
\begin{eqnarray}
w^{(4)}_{03}(1,1)&=& -\frac{2529Z_1a^3}{88\pi^2}\langle x^{-13}\rangle_{n_0L_0}
+\frac{1701a^3(3L_0^2+3L_0+44)}{88\pi^2}\langle x^{-14}\rangle_{n_0L_0}\,,
\label{eq213d}
\end{eqnarray}
\begin{eqnarray}
w^{(4)}_{11}(1,1)&=& -\frac{Z_1a^2}{4\pi^2}\langle x^{-11}\rangle_{n_0L_0}
+\frac{a^2(2L_0^2+2L_0+315)}{8\pi^2}\langle x^{-12}\rangle_{n_0L_0}\,,
\label{eq214}
\end{eqnarray}
\begin{eqnarray}
w^{(4)}_{11}(1,2)&=& -\frac{45Z_1a^2}{88\pi^2}\langle x^{-13}\rangle_{n_0L_0}
+\frac{45a^2(L_0^2+L_0+297)}{88\pi^2}\langle x^{-14}\rangle_{n_0L_0}\,,
\label{eq215}
\end{eqnarray}
\begin{eqnarray}
w^{(4)}_{11}(2,2)&=& -\frac{225Z_1a^2}{208\pi^2}\langle x^{-15}\rangle_{n_0L_0}
+\frac{225a^2(2L_0^2+2L_0+1001)}{416\pi^2}\langle x^{-16}\rangle_{n_0L_0}\,,
\label{eq216}
\end{eqnarray}
\begin{eqnarray}
w^{(4)}_{11}(1,3)&=& -\frac{21Z_1a^2}{26\pi^2}\langle x^{-15}\rangle_{n_0L_0}
+\frac{21a^2(2L_0^2+2L_0+1001)}{52\pi^2}\langle x^{-16}\rangle_{n_0L_0}\,.
\label{eq217}
\end{eqnarray}

Finally, we evaluate $W_g(2,k;L_{c_1},L_{c_3})$:
\begin{eqnarray}
W_g(2,k;L_{c_1},L_{c_3})&=& \sum_{M_{c_1}}\langle \chi_{n_0M_0L_0}|u_{L_{c_1}M_{c_1}}h^2_s
u^*_{L_{c_1}M_{c_1}}|g_k(L_{c_3})\rangle\nonumber\\
&=&\sum_{M_{c_1}}\langle \chi_{n_0M_0L_0}|[u_{L_{c_1}M_{c_1}},h_s][h_s,u^*_{L_{c_1}M_{c_1}}]|g_k(L_{c_3})\rangle\nonumber\\
&+&\sum_{M_{c_1}}\langle \chi_{n_0M_0L_0}|u_{L_{c_1}M_{c_1}}h_s
u^*_{L_{c_1}M_{c_1}}h_s|g_k(L_{c_3})\rangle\,.
\label{eq218}
\end{eqnarray}
In the above expression, the first term on the right-hand-side can be neglected because it contributes terms of order
$\langle x^{-11}\rangle_{n_0L_0}$ and below. The second term can be simplified by applying (\ref{eq196})
\begin{eqnarray}
W_g(2,k;L_{c_1},L_{c_3})&\approx&  \langle \chi_{n_0M_0L_0} |\mathcal{\hat{U}}_1(L_{c_1})\mathcal{\hat{U}}_k(L_{c_3})|
\chi_{n_0M_0L_0}\rangle \nonumber\\
&-&\langle \chi_{n_0M_0L_0} |\mathcal{\hat{U}}_1(L_{c_1})|\chi_{n_0M_0L_0}\rangle
\langle \chi_{n_0M_0L_0} |\mathcal{\hat{U}}_k(L_{c_3})|\chi_{n_0M_0L_0}\rangle\nonumber\\
&=&w^{(4)}_{1k}(L_{c_1},L_{c_3})-w^{(2)}_1(L_{c_1})w^{(2)}_k(L_{c_3})\,,
\label{eq219}
\end{eqnarray}
where (\ref{eq37}) and (\ref{eq207}) have been used. Therefore the final result for $\Delta E^{(1)}_{4b}$, accurate to
$\langle x^{-10}\rangle_{n_0L_0}$, is
\begin{eqnarray}
\Delta E^{(1)}_{4b} &=&-\frac{1}{4}\,q_x^4\,a^8\,\alpha_1^2S_{2,2}(n_0,L_0)-\frac{1}{2}\,q_x^4\,a^9\,\alpha_1(a\,\alpha_2-6\beta_1)S_{4,2}(n_0,L_0)\nonumber\\
&+&\frac{1}{2}\,q_x^4\,a^8\,\alpha_1\beta_1\langle x^{-8}\rangle_{n_0L_0}-\frac{1}{2}\,q_x^4\,a^8\,\alpha_1\beta_1 (\langle x^{-4}\rangle_{n_0L_0})^2\nonumber\\
&+&\frac{1}{2}\,q_x^4\,a^9\,(a\,\alpha_1\beta_2+a\,\alpha_2\beta_1-28\alpha_1\gamma_1-10\beta_1^2)\langle x^{-10}\rangle_{n_0L_0}\nonumber\\
&-&\frac{1}{2}\,q_x^4\,a^9\,(a\,\alpha_1\beta_2+a\,\alpha_2\beta_1-12\alpha_1\gamma_1-6\beta_1^2)\langle x^{-4}\rangle_{n_0L_0}\langle x^{-6}\rangle_{n_0L_0}+O(x^{-11})\,.
\label{eq220}
\end{eqnarray}
In the above, we have neglected $S_{4,4}(n_0,L_0)$ and $S_{6,2}(n_0,L_0)$.
Substituting (\ref{eq220}) and (\ref{eq186}) into (\ref{eq161}) and (\ref{eq151}), we finally arrive at the following expression for the total fourth-order energy correction
\begin{eqnarray}
\Delta E_4&=& \Delta E_{4a}^{(1)}-\frac{1}{4}\,q_x^4\,a^8\,\alpha_1^2 S_{2,2}(n_0,L_0)
-\frac{1}{2}\,q_x^4\,a^9\,\alpha_1(a\,\alpha_2-6\beta_1)S_{4,2}(n_0,L_0)\nonumber\\
&+&\frac{1}{2}\,q_x^4\,a^8\,\alpha_1\beta_1\langle x^{-8}\rangle_{n_0L_0}+
\frac{1}{2}\,q_x^4\,a^9\,(a\,\alpha_1\beta_2+a\,\alpha_2\beta_1-28\alpha_1\gamma_1-10\beta_1^2)\langle x^{-10}\rangle_{n_0L_0}\,,
\label{eq221}
\end{eqnarray}
where $\Delta E_{4a}^{(1)}$ is given by (\ref{eq169}). For the fifth-order correction $\Delta E_5$, it contributes terms of $O(x^{-11})$ and smaller and can thus be neglected.

%\subsection{The integral related to the reduced Schr\"{o}dinger-Coulomb Green function}

Finally let us discuss a scaling property of $S_{i,j}(n_0,L_0)$ defined in (\ref{eq192}), which was calculated by Swainson and Drake in \cite{Swainson} using $h'_x=-\nabla_r^2/2-1/r$ as the Rydberg electron Hamiltonian. It can also be calculated using equation (6.1.12) in \cite{Swainson2} where an extra factor of 2 needs to be applied because of the units used.
Our Hamiltonian $h_x=a(-\nabla_x^2/2-Z_1/x)$, however, can be transformed into $h'_x$ by letting $r=Z_1x$, {\it i.e.}, $h_x=aZ_1^2h'_x$. Since
\begin{eqnarray}
S_{i,j}(n_0,L_0)&=&\sum_{n_{x_2}L_{x_2}M_{x_2}} \langle \chi_{n_0 L_0 M_0} | x^{-i-2} \frac{1}{{e_{n_{x_2}}-e_{n_0}}} |\chi_{ n_{x_2}L_{x_2}M_{x_2}}\rangle\langle \chi_{n_{x_2}L_{x_2}M_{x_2}}| x^{-j-2}|\chi_{n_0 L_0 M_0}\rangle \nonumber
\label{eq221_scaling}
\end{eqnarray}
by applying the definition of $\hat{G}(n_0)$ in (\ref{eq189}), we then have
the corresponding transformation $S_{i,j}(n_0,L_0)=(Z_1^{i+j+2}/a)\,S'_{i,j}(n_0,L_0)$, where $S'_{i,j}(n_0,L_0)$ is the one calculated in \cite{Swainson}.

\section{Results and discussion}

After collecting all terms up to $\langle x^{-10}\rangle_{n_0L_0}$, the second-order correction $\triangle E_2$ can be expressed as follows
\begin{eqnarray}
\frac{\triangle E_2}{q_x^2} &=& \sum_{L_c=1}^4 \left(-\frac{1}{2}\right)a^{2L_c+2}\alpha_{L_c}\langle x^{-2L_c-2}\rangle_{n_0L_0} \nonumber\\
&+& \sum_{L_c=1}^3 \frac{1}{2}a^{2L_c+3}(L_c+1)(2L_c+1)\beta_{L_c}\langle x^{-2L_c-4}\rangle_{n_0L_0} \nonumber\\
&+& \sum_{L_c=1}^2 a^{2L_c+4}\gamma_{L_c}(L_c+1)^2\left\{\frac{2Z_1}{2L_c+3}\langle x^{-2L_c-5}\rangle_{n_0L_0} -(L_c+2)(2L_c+1)\left[1+\frac{L_0(L_0+1)}{(L_c+1)(2L_c+3)}\right]\right. \nonumber\\
& &\times \langle x^{-2L_c-6}\rangle_{n_0L_0}\bigg\} \nonumber\\
&+& a^7 \delta_1 \left\{ -\frac{408Z_1}{7}\langle x^{-9}\rangle_{n_0L_0} + 720\left[1+\frac{3}{14}L_0(L_0+1)\right]\langle x^{-10}\rangle_{n_0L_0} \right\} \nonumber\\
&+& a^8 \varsigma_1\left(-\frac{164Z_1^2}{7}\langle x^{-10}\rangle_{n_0L_0}\right)\,,
\label{eqe2}
\end{eqnarray}
where we have moved the $q_x$-related factor to the left hand side. It should be noted that the last term in the above expression of $\triangle E_2$
is absent in both Drake's \cite{Drake_Adv} and Drachman's \cite{DrachmanHe} formulas.

The third-order correction $\triangle E_3$ reads
\begin{eqnarray}
\frac{\triangle E_3}{-q_x^3} &=& -a^7 \pi^{\frac{3}{2}}\left(\frac{16\sqrt{10}}{225}w^{(3)}_c(0,0;1,1,2)+\frac{8\sqrt{6}}{135}w^{(3)}_c(0,0;1,2,1)\right)\langle x^{-7}\rangle_{n_0L_0} \nonumber\\
&+& a^9\pi^{\frac{3}{2}}\left(\frac{16\sqrt{21}}{735}w^{(3)}_c(0,0;1,2,3)+\frac{16\sqrt{15}}{525}w^{(3)}_c(0,0;1,3,2)+\frac{16\sqrt{35}}{1225}w^{(3)}_c(0,0;2,1,3)\right.\nonumber\\
& &+\left.\frac{8\sqrt{14}}{875}w^{(3)}_c(0,0;2,2,2)\right)\langle x^{-9}\rangle_{n_0L_0}\nonumber\\
&+& 2a^8\pi^{\frac{3}{2}}\left(\frac{8\sqrt{10}}{25}w^{(3)}_c(1,0;1,1,2)+\frac{16\sqrt{6}}{45}w^{(3)}_c(1,0;1,2,1)+\frac{16\sqrt{10}}{75}w^{(3)}_c(1,0;2,1,1)\right)\langle x^{-9}\rangle_{n_0L_0}\nonumber\\
&+& a^9\pi^{\frac{3}{2}}Z_1\left(\frac{4\sqrt{10}}{75}w^{(3)}_c(1,1;1,1,2)+\frac{4\sqrt{6}}{135}w^{(3)}_c(1,1;1,2,1)\right)\langle x^{-10}\rangle_{n_0L_0}\nonumber\\
&+& 2a^9\pi^{\frac{3}{2}}Z_1\left(\frac{4\sqrt{10}}{75}w^{(3)}_c(2,0;1,1,2)+\frac{2\sqrt{6}}{27}w^{(3)}_c(2,0;1,2,1)+\frac{2\sqrt{10}}{45}w^{(3)}_c(2,0;2,1,1)\right)\langle x^{-10}\rangle_{n_0L_0}\,.\nonumber\\
\label{eqe3}
\end{eqnarray}
It should be noted again that all the $\langle x^{-10}\rangle_{n_0L_0}$ terms above are entirely missing in the works of Drake~\cite{Drake_Adv} and Drachman~\cite{DrachmanHe}.

Finally the expression for $\triangle E_4$ is
\begin{eqnarray}
\frac{\triangle E_4}{q_x^4} &=& -a^8\pi ^2\left(\frac{16}{81}w^{(4)}_c(0,0,0;0,1,1,1,1)+\frac{32}{405}w^{(4)}_c(0,0,0;2,1,1,1,1)\right)\langle x^{-8}\rangle_{n_0L_0} \nonumber\\
&+& 2a^9\pi ^2\left(\frac{112}{81}w^{(4)}_c(0,0,1;0,1,1,1,1)+\frac{224}{405}w^{(4)}_c(0,0,1;2,1,1,1,1)\right)\langle x^{-10}\rangle_{n_0L_0} \nonumber\\
&+& a^9\pi ^2\left(\frac{128}{81}w^{(4)}_c(0,1,0;0,1,1,1,1)+\frac{352}{405}w^{(4)}_c(0,1,0;2,1,1,1,1)\right)\langle x^{-10}\rangle_{n_0L_0} \nonumber\\
&+& a^{10}\pi^2\left(\frac{32\sqrt{6}}{945}w^{(4)}_c(0,0,0;2,1,1,3,1)+\frac{32}{675}w^{(4)}_c(0,0,0;1,1,2,2,1)+\frac{16}{525}w^{(4)}_c(0,0,0;3,1,2,2,1)\right. \nonumber\\
&+& \frac{32}{225}w^{(4)}_c(0,0,0;0,1,1,2,2)+\frac{64\sqrt{35}}{7875}w^{(4)}_c(0,0,0;2,1,1,2,2)+\frac{64\sqrt{15}}{3375}w^{(4)}_c(0,0,0;1,2,1,2,1) \nonumber\\
&+& \frac{32\sqrt{15}}{2625}w^{(4)}_c(0,0,0;3,2,1,2,1)+\frac{32\sqrt{14}}{2205}w^{(4)}_c(0,0,0;2,1,1,1,3)+\frac{32}{1125}w^{(4)}_c(0,0,0;1,2,1,1,2) \nonumber\\
&+& \frac{16}{875}w^{(4)}_c(0,0,0;3,2,1,1,2)\Bigg)\langle x^{-10}\rangle_{n_0L_0} \nonumber\\
&+& \frac{1}{2}a^8\alpha_1\beta _1\langle x^{-8}\rangle_{n_0L_0}+\frac{1}{2}a^9\left(a \alpha _1\beta _2+a \alpha _2\beta _1-28\alpha _1\gamma _1-10\beta_1^2\right)\langle x^{-10}\rangle_{n_0L_0} \nonumber\\
&-& \frac{1}{4}a^8\alpha_1^2S_{2,2}\left(n_0,L_0\right)-\frac{1}{2}a^9\alpha _1\left(a \alpha _2-6\beta _1\right)S_{2,4}\left(n_0,L_0\right)\,.
\label{eqe4}
\end{eqnarray}
The above expression is in agreement with Drake's formula~\cite{Drake_Adv} and differs from the result of Drachman~\cite{DrachmanHe} regarding the term
$(-28\alpha _1\gamma _1-10\beta_1^2)\langle x^{-10}\rangle_{n_0L_0}$. In Drachman's calculation, he obtains
$(-12\alpha _1\gamma _1-14\beta_1^2)\langle x^{-10}\rangle_{n_0L_0}$ instead.

The expressions in (\ref{eqe2}), (\ref{eqe3}), and (\ref{eqe4}) are valid for any atomic system in a high-$L$ atomic state with the core in an $S$-state as far as the nonrelativistic Hamiltonian (\ref{eq1}) is concerned.
For helium-like systems, all quantities of describing the core properties, such as $\alpha(i,L_c)$ in (\ref{eq40}), $w^{(3)}_c$ in (\ref{eq124}), and $w^{(4)}_c$ in (\ref{eq163}) can be calculated either analytically or numerically. For $\alpha(1,3)$, for example, our numerical result is $102.03125000000000(2) Z^{-10}$ using a 60-term Sturmian basis set~\cite{yan_sturmian}, while the analytical value given in \cite{Drake_Adv} is $\frac{3265}{32} Z^{-10}$. We have checked the analytical values listed in \cite{Drake_Adv} and contained in \cite{DrachmanHe} and found that all are correct except $\theta$, the nonadiabatic correction to the term $\langle x^{-8}\rangle_{n_0L_0}$ in (\ref{eqe4}), {\it i.e.},
\begin{eqnarray}
\theta&=&2\bigg[
2a^9\pi ^2\left(\frac{112}{81}w^{(4)}_c(0,0,1;0,1,1,1,1)+\frac{224}{405}w^{(4)}_c(0,0,1;2,1,1,1,1)\right)\langle x^{-10}\rangle_{n_0L_0} \nonumber\\
&+& a^9\pi ^2\left(\frac{128}{81}w^{(4)}_c(0,1,0;0,1,1,1,1)+\frac{352}{405}w^{(4)}_c(0,1,0;2,1,1,1,1)\right)\langle x^{-10}\rangle_{n_0L_0}\bigg]\,.
\end{eqnarray}
The value $\frac{791313}{128}Z^{-12}$ of $\theta$ used in \cite{Drake_Adv} and \cite{DrachmanHe} is incorrect and it should be $8348.7968750000000000(1)Z^{-12}$ numerically. To verify this, we carried out an analytical derivation using a method similar to \cite{DrachmanHe} and obtained $\theta=\frac{534323}{64}Z^{-12}$ that is in agreement with our numerical value.

The finite nuclear mass effect is fully considered in our derivation of expressions $\Delta E_2$, $\Delta E_3$, and $\Delta E_4$ either explicitly
through the parameter $a=\mu_x/\mu$ or implicitly through the nuclear mass related parameters, such as in
$T_{\ell m}({\bf r}_1,{\bf r}_2,\ldots,{\bf r}_n)$ defined by (\ref{eq13}). It is possible to express the total energy as a sum of the $0$th-order energy and a series expansion of corrections in powers of $y=\mu/M$.
For a helium-like system, we have
\begin{eqnarray}
E_M=-\frac{Z^2}{2}-\frac{(Z-1)^2}{2n_0^2}+\Delta E_\infty+y\varepsilon_M^{(1)}+y^2\varepsilon_M^{(2)}+y^3\varepsilon_M^{(3)}+y^4\varepsilon_M^{(4)}
+\cdots\,,
\label{eqeinfEM}
\end{eqnarray}
in $2R_M$. In the above,
\begin{eqnarray}
\Delta E_{\infty} &=& -\frac{9}{4 Z^4}\langle x^{-4}\rangle_{n_0L_0}
+ \frac{69}{8 Z^6} \langle x^{-6}\rangle_{n_0L_0}
+\frac{319}{30 Z^8}\left(Z+\frac{2557}{638}\right) \langle x^{-7}\rangle_{n_0L_0} \nonumber\\
&-&\frac{957}{40 Z^{10}} \left[Z^2 L_0(L_0+1)+\frac{5455}{638}Z^2+\frac{5925}{2552}\right]\langle x^{-8}\rangle_{n_0L_0} \nonumber\\
&-&\frac{1}{672 Z^{10}}\left(307291Z+293603\right)\langle x^{-9}\rangle_{n_0L_0} \nonumber\\
&+&\frac{1}{48384 Z^{12}}[57499200 Z^2L_0(L_0+1)-12199181 Z^4+24398362 Z^3 \nonumber\\
& &+189168979 Z^2 -53398422 Z+23252544] \langle x^{-10}\rangle_{n_0L_0} \nonumber\\
&-&\frac{81(Z-1)^6}{16 Z^8}  S'_{2,2}(n_0,L_0)
+\frac{621(Z-1)^8}{16 Z^{10}}  S'_{2,4}(n_0,L_0)\,,
\label{eqeinf}
\end{eqnarray}
\begin{eqnarray}
\varepsilon_M^{(1)} &=& -\frac{9}{2 Z^4} (Z-1) \langle x^{-4}\rangle_{n_0L_0}
+ \frac{3}{4 Z^6} (43 Z-3) \langle x^{-6}\rangle_{n_0L_0}
+ \frac{1}{15 Z^8}\left(319Z^2+\frac{1919}{2}Z-2876\right) \langle x^{-7}\rangle_{n_0L_0} \nonumber\\
&-& \frac{1}{Z^{10}}\left[\frac{957}{20}Z^2(Z-1)L_0(L_0+1)+\frac{957}{2} Z^3-471 Z^2+\frac{3555}{16} Z-\frac{3555}{16}\right] \langle x^{-8}\rangle_{n_0L_0} \nonumber\\
&-& \frac{1}{168 Z^{10}}\left(164441 Z^2+154940 Z-160180\right) \langle x^{-9}\rangle_{n_0L_0} \nonumber\\
&+& \frac{1}{8Z^{12}} \left[\frac{5}{7} Z^2 (29019 Z-24221)L_0(L_0+1)-\frac{12199181}{3024} Z^5+\frac{12199181}{1008} Z^4\right. \nonumber\\
& &+\left.\frac{85304659}{1008} Z^3-\frac{117661081}{3024} Z^2+\frac{727807}{8} Z+\frac{157111}{3}\right] \langle x^{-10}\rangle_{n_0L_0} \nonumber\\
&-& \frac{81}{4 Z^8} (Z-1)^7 S'_{2,2}(n_0,L_0)
+\frac{27}{4 Z^{10}} (33 Z-13) (Z-1)^8 S'_{2,4}(n_0,L_0)\,,
\label{eqm1}
\end{eqnarray}
\begin{eqnarray}
\varepsilon_M^{(2)}&=&-\frac{(Z-1)^2}{2n_0^2}+\tilde{\varepsilon}_M^{(2)}
\end{eqnarray}
with
\begin{eqnarray}
\tilde{\varepsilon}_M^{(2)} &=& -\frac{9}{4 Z^4} \left(Z^2-2 Z+5\right) \langle x^{-4}\rangle_{n_0L_0}
+\frac{3}{8 Z^6} \left(43 Z^2-46 Z+18\right)\langle x^{-6}\rangle_{n_0L_0} \nonumber\\
&+& \frac{1}{30 Z^8}\left(319 Z^3+\frac{1281}{2} Z^2-\frac{16623}{2} Z+\frac{37069}{2}\right) \langle x^{-7}\rangle_{n_0L_0} \nonumber\\
&-& \frac{1}{4Z^{10}} \left[\frac{957}{10} Z^2 \left(Z^2-2 Z+7\right)L_0(L_0+1) + 957 Z^4-\frac{2223}{2} Z^3\right. \nonumber\\
& & \left.+\frac{70827}{8} Z^2-\frac{10665}{4} Z+\frac{24885}{8}\right]\langle x^{-8}\rangle_{n_0L_0} \nonumber\\
&-& \frac{1}{336 Z^{10}} \left(164441 Z^3+126579 Z^2-1006423 Z+63227\right) \langle x^{-9}\rangle_{n_0L_0} \nonumber\\
&+& \frac{1}{48384 Z^{12}}[2160 Z^2 \left(29019 Z^2-53240 Z+198566\right) L_0(L_0+1) -12199181 Z^6+48796724 Z^5 \nonumber\\
& & +121722986 Z^4-176800562 Z^3+1562925159 Z^2-635022108 Z-1699941600]\langle x^{-10}\rangle_{n_0L_0} \nonumber\\
&-& \frac{81 }{16 Z^8}\left(6 Z^2-12 Z+13\right) (Z-1)^6 S'_{2,2}(n_0,L_0)
+\frac{27 }{8 Z^{10}} \left(119 Z^2-138 Z+61\right) (Z-1)^8 S'_{2,4}(n_0,L_0)\,,\nonumber\\
\label{eqm2}
\end{eqnarray}
\begin{eqnarray}
\varepsilon_M^{(3)} &=& -\frac{18}{Z^4}(Z-1) \langle x^{-4}\rangle_{n_0L_0}
+\frac{15 }{4 Z^6}(35 Z+13) \langle x^{-6}\rangle_{n_0L_0} \nonumber\\
&-&\frac{427}{5 Z^8}\left(Z^2-\frac{1309}{122} Z+\frac{1126}{61}\right) \langle x^{-7}\rangle_{n_0L_0} + O\left(\langle x^{-8}\rangle_{n_0L_0}\right)\,,
\label{eqm3}
\end{eqnarray}
\begin{eqnarray}
\varepsilon_M^{(4)}&=&-\frac{(Z-1)^2}{2n_0^2}+\tilde{\varepsilon}_M^{(4)}
\end{eqnarray}
with
\begin{eqnarray}
\tilde{\varepsilon}_M^{(4)} &=& -\frac{9}{2 Z^4} \left(2 Z^2-4 Z+7\right)\langle x^{-4}\rangle_{n_0L_0}
+\frac{45}{8 Z^6} \left(13 Z^2-10 Z-20\right) \langle x^{-6}\rangle_{n_0L_0} \nonumber\\
&+&\frac{1}{20 Z^8} \left(211Z^3+6822 Z^2-47086 Z+69873\right)\langle x^{-7}\rangle_{n_0L_0} + O\left(\langle x^{-8}\rangle_{n_0L_0}\right)\,.
\label{eqm4}
\end{eqnarray}

Tables I to VI in the Supplemental Material \cite{Supplemental_Material} list numerical values for $\Delta E_{\infty}$, $\varepsilon_M^{(1)}$, and $\tilde{\varepsilon}_M^{(2)}$ of helium in Rydberg states with $L_0$ from 4 to 15 and $n_0$ from $L_0+1$ to $16$, where $\Delta_n$ ($n=4,6,7,8,9,10$) denotes the contribution of the terms involving $\langle x^{-n} \rangle_{n_0L_0}$, and $\Delta_{2,2}$ and $\Delta_{2,4}$ denote, respectively, the contributions involving $S'_{2,2}\left(n_0,L_0\right)$ and $S'_{2,4}\left(n_0,L_0\right)$. In these tables, we keep $10$ significant figures for all the numbers. Our results could serve as a benchmark for future reference.

In summary, we have presented a complete calculation for the nonrelativistic energy levels of a Rydberg atom up to the order of $\langle x^{-10}\rangle_{n_0L_0}$. We have also corrected the existing errors in the literature and recovered various missing terms from the previous works.
It is desirable to revisit relativistic and quantum electrodynamic corrections~\cite{DrakePlenum,Drake_Adv,DrachmanHe} to the nonrelativistic energies so that a meaningful comparison
with experimental measurements can be made. Work along this direction
is currently underway.

\begin{acknowledgements}
XFW wishes to thank the China Scholarship Council for supporting his research visit to the Department of Physics of the University of New Brunswick from December 2014 to November 2015.
ZCY was supported by NSERC of Canada and by the CAS/SAFEA International Partnership Program for Creative Research Teams.
Research support from the computing facilities of SHARCnet and ACEnet is gratefully acknowledged.
\end{acknowledgements}

\end{document}

\title{\bf Supplemental Material for "General theory for Rydberg states of atoms: nonrelativistic case"}
\author{Xiao-Feng Wang$^{1,2}$, Zong-Chao Yan$^{2,3,4}$}

\affiliation{$^1$ School of Physics and Technology, Wuhan University, Wuhan 430072, P. R. China}

\affiliation{$^2$ Department of Physics, University of New
Brunswick, Fredericton, New Brunswick, Canada E3B 5A3}

\affiliation {$^3$ State Key Laboratory of Magnetic Resonance and
Atomic and Molecular Physics, Wuhan Institute of Physics and
Mathematics, Chinese Academy of Sciences, Wuhan 430071, P. R. China}

\affiliation {$^4$ Center for Cold Atom Physics, Chinese Academy of
Sciences, Wuhan 430071, P. R. China}

\date{\today}

 \maketitle

\clearpage

\begin{turnpage}
\begin{table}
\renewcommand\arraystretch{0.8}
\caption{$\Delta E_{\infty}$ and its components for $L_0=4$ to $7$ and $n_0=L_0+1$ to $16$, in $2R_{M}$. Numbers in square brackets denote powers of $10$.}
\label{tab1}
{\tiny
\begin{ruledtabular}
\begin{tabular}{r r r r r r r r r r r}
\multicolumn{1}{c}{$L_0$} & $n_0$ & \multicolumn{1}{c}{$\Delta_4$} & \multicolumn{1}{c}{$\Delta_6$} & \multicolumn{1}{c}{$\Delta_{7}$} & \multicolumn{1}{c}{$\Delta_{8}$} & \multicolumn{1}{c}{$\Delta_{9}$} & \multicolumn{1}{c}{$\Delta_{10}$} & \multicolumn{1}{c}{$\Delta_{2,2}$} & \multicolumn{1}{c}{$\Delta_{2,4}$} & \multicolumn{1}{c}{$\Delta E_{\infty}$} \\
\hline
$4$ & $5$ & $-7.142857143[-7]$ & $3.650793651[-9]$ & $6.760141093[-10]$ & $-9.833498677[-10]$ & $-9.534150080[-11]$ & $7.688444846[-10]$ & $-1.453423631[-10]$ & $3.517562448[-12]$ & $-7.104105782[-7]$ \\

$4$ & $6$ & $-4.592886537[-7]$ & $3.056608517[-9]$ & $6.288771040[-10]$ & $-9.910149199[-10]$ & $-1.016280188[-10]$ & $8.485132903[-10]$ & $-1.096301780[-10]$ & $3.373368731[-12]$ & $-4.559535546[-7]$ \\

$4$ & $7$ & $-3.066710763[-7]$ & $2.298376708[-9]$ & $4.955480268[-10]$ & $-8.108352933[-10]$ & $-8.551905847[-11]$ & $7.273307632[-10]$ & $-7.860952282[-11]$ & $2.695926167[-12]$ & $-3.041220887[-7]$ \\

$4$ & $8$ & $-2.130285486[-7]$ & $1.706215210[-9]$ & $3.777523205[-10]$ & $-6.317048521[-10]$ & $-6.774618155[-11]$ & $5.826901672[-10]$ & $-5.671835483[-11]$ & $2.070103524[-12]$ & $-2.111159902[-7]$ \\

$4$ & $9$ & $-1.532680431[-7]$ & $1.279795568[-9]$ & $2.881476157[-10]$ & $-4.886488546[-10]$ & $-5.297575680[-11]$ & $4.590404437[-10]$ & $-4.172787664[-11]$ & $1.585379723[-12]$ & $-1.518228266[-7]$ \\

$4$ & $10$ & $-1.136363636[-7]$ & $9.759251166[-10]$ & $2.222603207[-10]$ & $-3.805484478[-10]$ & $-4.156671187[-11]$ & $3.620471414[-10]$ & $-3.137346863[-11]$ & $1.225623050[-12]$ & $-1.125283941[-7]$ \\

$4$ & $11$ & $-8.643507054[-8]$ & $7.572957345[-10]$ & $1.738869661[-10]$ & $-2.997847390[-10]$ & $-3.292284576[-11]$ & $2.878384960[-10]$ & $-2.408297904[-11]$ & $9.600896504[-13]$ & $-8.557187982[-8]$ \\
\end{tabular}
\end{ruledtabular}
}
\end{table}
\end{turnpage}

%table for $\Delta E_{\infty}$
\begin{turnpage}
\begin{table}
\renewcommand\arraystretch{0.8}
\caption{$\Delta E_{\infty}$ and its components for $L_0=8$ to $15$ and $n_0=L_0+1$ to $16$, in $2R_{M}$. Numbers in square brackets denote powers of $10$.}
\label{tab2}
{\tiny
\begin{ruledtabular}
\begin{tabular}{r r r r r r r r r r r}
\multicolumn{1}{c}{$L_0$} & $n_0$ & \multicolumn{1}{c}{$\Delta_4$} & \multicolumn{1}{c}{$\Delta_6$} & \multicolumn{1}{c}{$\Delta_{7}$} & \multicolumn{1}{c}{$\Delta_{8}$} & \multicolumn{1}{c}{$\Delta_{9}$} & \multicolumn{1}{c}{$\Delta_{10}$} & \multicolumn{1}{c}{$\Delta_{2,2}$} & \multicolumn{1}{c}{$\Delta_{2,4}$} & \multicolumn{1}{c}{$\Delta E_{\infty}$} \\
\hline
$8$ & $9$ & $-4.669601418[-9]$ & $1.214227294[-12]$ & $4.163654695[-14]$ & $-2.555750680[-14]$ & $-9.885860631[-17]$ & $1.608240692[-16]$ & $-1.650665866[-14]$ & $1.474227633[-17]$ & $-4.668387541[-9]$ \\

$8$ & $10$ & $-3.676470588[-9]$ & $1.226051318[-12]$ & $4.779511520[-14]$ & $-3.300500593[-14]$ & $-1.417094435[-16]$ & $2.523414851[-16]$ & $-1.486748497[-14]$ & $1.716560515[-17]$ & $-3.675244486[-9]$ \\

$8$ & $11$ & $-2.913572385[-9]$ & $1.117147917[-12]$ & $4.650050889[-14]$ & $-3.410931822[-14]$ & $-1.547347566[-16]$ & $2.895382210[-16]$ & $-1.274744681[-14]$ & $1.687495506[-17]$ & $-2.912455441[-9]$ \\

$8$ & $12$ & $-2.332880643[-9]$ & $9.781656812[-13]$ & $4.241485141[-14]$ & $-3.231961903[-14]$ & $-1.518737286[-16]$ & $2.935043214[-16]$ & $-1.073584887[-14]$ & $1.549813521[-17]$ & $-2.331902960[-9]$ \\

$8$ & $13$ & $-1.889160013[-9]$ & $8.427263880[-13]$ & $3.758354329[-14]$ & $-2.940359934[-14]$ & $-1.416151020[-16]$ & $2.799755664[-16]$ & $-8.998281876[-15]$ & $1.379428615[-17]$ & $-1.888317952[-9]$ \\

$8$ & $14$ & $-1.547055066[-9]$ & $7.220163852[-13]$ & $3.286651148[-14]$ & $-2.621500857[-14]$ & $-1.285689040[-16]$ & $2.585015144[-16]$ & $-7.551083678[-15]$ & $1.209791807[-17]$ & $-1.546333807[-9]$ \\

$8$ & $15$ & $-1.280434201[-9]$ & $6.183902653[-13]$ & $2.859016124[-14]$ & $-2.314216060[-14]$ & $-1.150822871[-16]$ & $2.343940413[-16]$ & $-6.362578400[-15]$ & $1.054320541[-17]$ & $-1.279816595[-9]$ \\
\end{tabular}
\end{ruledtabular}
}
\end{table}
\end{turnpage}

%table for $\varepsilon_M^{(1)}$
\begin{turnpage}
\begin{table}
\renewcommand\arraystretch{0.8}
\caption{$\varepsilon_M^{(1)}$ and its components for $L_0=4$ to $7$ and $n_0=L_0+1$ to $16$, in $2R_{M}$. Numbers in square brackets denote powers of $10$.}
\label{tab3}
{\tiny
\begin{ruledtabular}
\begin{tabular}{r r r r r r r r r r r}
\multicolumn{1}{c}{$L_0$} & $n_0$ & \multicolumn{1}{c}{$\Delta_4$} & \multicolumn{1}{c}{$\Delta_6$} & \multicolumn{1}{c}{$\Delta_{7}$} & \multicolumn{1}{c}{$\Delta_{8}$} & \multicolumn{1}{c}{$\Delta_{9}$} & \multicolumn{1}{c}{$\Delta_{10}$} & \multicolumn{1}{c}{$\Delta_{2,2}$} & \multicolumn{1}{c}{$\Delta_{2,4}$} & \multicolumn{1}{c}{$\varepsilon_M^{(1)}$} \\
\hline
$4$ & $5$ & $-1.428571429[-6]$ & $2.634920635[-8]$ & $2.250440917[-10]$ & $-2.114351852[-9]$ & $-3.390711346[-10]$ & $2.427383393[-9]$ & $-5.813694525[-10]$ & $3.242274952[-11]$ & $-1.402572164[-6]$ \\

$4$ & $6$ & $-9.185773075[-7]$ & $2.206073973[-8]$ & $2.093522527[-10]$ & $-2.130832880[-9]$ & $-3.614284162[-10]$ & $2.67891[-9]$ & $-4.385207120[-10]$ & $3.109365961[-11]$ & $-8.965279914[-7]$ \\

$4$ & $7$ & $-6.133421526[-7]$ & $1.658828406[-8]$ & $1.649672012[-10]$ & $-1.743419265[-9]$ & $-3.041387428[-10]$ & $2.296316942[-9]$ & $-3.144380913[-10]$ & $2.484940641[-11]$ & $-5.966297311[-7]$ \\

$4$ & $8$ & $-4.260570972[-7]$ & $1.231442282[-8]$ & $1.257531857[-10]$ & $-1.358261558[-9]$ & $-2.409315403[-10]$ & $1.839659988[-9]$ & $-2.268734193[-10]$ & $1.908095422[-11]$ & $-4.135842468[-7]$ \\

$4$ & $9$ & $-3.065360861[-7]$ & $9.236785403[-9]$ & $9.592391278[-11]$ & $-1.050669395[-9]$ & $-1.884022153[-10]$ & $1.449275077[-9]$ & $-1.669115065[-10]$ & $1.461306528[-11]$ & $-2.971454718[-7]$ \\

$4$ & $10$ & $-2.272727273[-7]$ & $7.043633450[-9]$ & $7.399013024[-11]$ & $-8.182370707[-10]$ & $-1.478272530[-10]$ & $1.143049389[-9]$ & $-1.254938745[-10]$ & $1.129704725[-11]$ & $-2.200923155[-7]$ \\

$4$ & $11$ & $-1.728701411[-7]$ & $5.465699649[-9]$ & $5.788671242[-11]$ & $-6.445828070[-10]$ & $-1.170863325[-10]$ & $9.087590520[-10]$ & $-9.633191615[-11]$ & $8.849521995[-12]$ & $-1.672869472[-7]$ \\
\end{tabular}
\end{ruledtabular}
}
\end{table}
\end{turnpage}

%table for $\varepsilon_M^{(1)}$
\begin{turnpage}
\begin{table}
\renewcommand\arraystretch{0.8}
\caption{$\varepsilon_M^{(1)}$ and its components for $L_0=8$ to $15$ and $n_0=L_0+1$ to $16$, in $2R_{M}$. Numbers in square brackets denote powers of $10$.}
\label{tab4}
{\tiny
\begin{ruledtabular}
\begin{tabular}{r r r r r r r r r r r}
\multicolumn{1}{c}{$L_0$} & $n_0$ & \multicolumn{1}{c}{$\Delta_4$} & \multicolumn{1}{c}{$\Delta_6$} & \multicolumn{1}{c}{$\Delta_{7}$} & \multicolumn{1}{c}{$\Delta_{8}$} & \multicolumn{1}{c}{$\Delta_{9}$} & \multicolumn{1}{c}{$\Delta_{10}$} & \multicolumn{1}{c}{$\Delta_{2,2}$} & \multicolumn{1}{c}{$\Delta_{2,4}$} & \multicolumn{1}{c}{$\varepsilon_M^{(1)}$} \\
\hline

$8$ & $13$ & $-3.778320025[-9]$ & $6.082286105[-12]$ & $1.251150567[-14]$ & $-6.039244480[-14]$ & $-5.036379008[-16]$ & $7.641517755[-16]$ & $-3.599312750[-14]$ & $1.271473332[-16]$ & $-3.772321225[-9]$ \\

$8$ & $14$ & $-3.094110131[-9]$ & $5.211074780[-12]$ & $1.094121280[-14]$ & $-5.384335568[-14]$ & $-4.572405908[-16]$ & $7.055415363[-16]$ & $-3.020433471[-14]$ & $1.115112448[-16]$ & $-3.088971803[-9]$ \\

$8$ & $15$ & $-2.560868402[-9]$ & $4.463164524[-12]$ & $9.517622161[-15]$ & $-4.753199227[-14]$ & $-4.092769815[-16]$ & $6.397437645[-16]$ & $-2.545031360[-14]$ & $9.718084991[-17]$ & $-2.556468374[-9]$ \\

$8$ & $16$ & $-2.140600246[-9]$ & $3.831574723[-12]$ & $8.270541685[-15]$ & $-4.178281370[-14]$ & $-3.637100980[-16]$ & $5.743304483[-16]$ & $-2.156114130[-14]$ & $8.454272290[-17]$ & $-2.136823449[-9]$ \\

$9$ & $10$ & $-3.869969040[-9]$ & $2.230607155[-12]$ & $2.721605211[-15]$ & $-9.545070875[-15]$ & $-3.942893009[-17]$ & $4.435527531[-17]$ & $-1.373375727[-14]$ & $1.700946604[-17]$ & $-3.867758968[-9]$ \\

$9$ & $11$ & $-3.123830186[-9]$ & $2.289308340[-12]$ & $3.174122293[-15]$ & $-1.254892966[-14]$ & $-5.776592574[-17]$ & $7.151276567[-17]$ & $-1.260001499[-14]$ & $2.007397569[-17]$ & $-3.121562818[-9]$ \\

$9$ & $12$ & $-2.532841840[-9]$ & $2.134359494[-12]$ & $3.164791940[-15]$ & $-1.331927051[-14]$ & $-6.495352184[-17]$ & $8.476955769[-17]$ & $-1.104520865[-14]$ & $2.020760485[-17]$ & $-2.530728641[-9]$ \\
\end{tabular}
\end{ruledtabular}
}
\end{table}
\end{turnpage}

%table for $\tilde{\varepsilon}_M^{(2)}$
\begin{turnpage}
\begin{table}
\renewcommand\arraystretch{0.8}
\caption{$\tilde{\varepsilon}_M^{(2)}$ and its components for $L_0=4$ to $7$ and $n_0=L_0+1$ to $16$, in $2R_{M}$. Numbers in square brackets denote powers of $10$.}
\label{tab5}
{\tiny
\begin{ruledtabular}
\begin{tabular}{r r r r r r r r r r r}
\multicolumn{1}{c}{$L_0$} & $n_0$ & \multicolumn{1}{c}{$\Delta_4$} & \multicolumn{1}{c}{$\Delta_6$} & \multicolumn{1}{c}{$\Delta_{7}$} & \multicolumn{1}{c}{$\Delta_{8}$} & \multicolumn{1}{c}{$\Delta_{9}$} & \multicolumn{1}{c}{$\Delta_{10}$} & \multicolumn{1}{c}{$\Delta_{2,2}$} & \multicolumn{1}{c}{$\Delta_{2,4}$} & \multicolumn{1}{c}{$\tilde{\varepsilon}_M^{(2)}$} \\
\hline
$4$ & $5$ & $-3.571428571[-6]$ & $1.555555556[-8]$ & $2.478130511[-9]$ & $-8.219014550[-9]$ & $2.682770639[-11]$ & $5.914765836[-9]$ & $-1.889450721[-9]$ & $7.983337382[-11]$ & $-3.557481924[-6]$ \\

$4$ & $6$ & $-2.296443269[-6]$ & $1.302381020[-8]$ & $2.305335817[-9]$ & $-8.283080431[-9]$ & $2.859664076[-11]$ & $6.527662644[-9]$ & $-1.425192314[-9]$ & $7.656080338[-11]$ & $-2.284189575[-6]$ \\

$4$ & $7$ & $-1.533355381[-6]$ & $9.793083364[-9]$ & $1.816578483[-9]$ & $-6.777106798[-9]$ & $2.406381452[-11]$ & $5.595398336[-9]$ & $-1.021923797[-9]$ & $6.118580258[-11]$ & $-1.523864102[-6]$ \\

$4$ & $8$ & $-1.065142743[-6]$ & $7.269960461[-9]$ & $1.384763333[-9]$ & $-5.279902445[-9]$ & $1.906278643[-11]$ & $4.482669725[-9]$ & $-7.373386128[-10]$ & $4.698234954[-11]$ & $-1.057956545[-6]$ \\

$4$ & $9$ & $-7.663402154[-7]$ & $5.453041985[-9]$ & $1.056290673[-9]$ & $-4.084214762[-9]$ & $1.490660455[-11]$ & $3.531425130[-9]$ & $-5.424623963[-10]$ & $3.598122677[-11]$ & $-7.608752469[-7]$ \\

$4$ & $10$ & $-5.681818182[-7]$ & $4.158289627[-9]$ & $8.147612226[-10]$ & $-3.180692175[-9]$ & $1.169626587[-11]$ & $2.785249951[-9]$ & $-4.078550922[-10]$ & $2.781631445[-11]$ & $-5.639725521[-7]$ \\

$4$ & $11$ & $-4.321753527[-7]$ & $3.226738347[-9]$ & $6.374343230[-10]$ & $-2.505654612[-9]$ & $9.264008142[-12]$ & $2.214358478[-9]$ & $-3.130787275[-10]$ & $2.178986076[-11]$ & $-4.288845010[-7]$ \\
\end{tabular}
\end{ruledtabular}
}
\end{table}
\end{turnpage}

%table for $\tilde{\varepsilon}_M^{(2)}$
\begin{turnpage}
\begin{table}
\renewcommand\arraystretch{0.8}
\caption{$\tilde{\varepsilon}_M^{(2)}$ and its components for $L_0=8$ to $15$ and $n_0=L_0+1$ to $16$, in $2R_{M}$. Numbers in square brackets denote powers of $10$.}
\label{tab6}
{\tiny
\begin{ruledtabular}
\begin{tabular}{r r r r r r r r r r r}
\multicolumn{1}{c}{$L_0$} & $n_0$ & \multicolumn{1}{c}{$\Delta_4$} & \multicolumn{1}{c}{$\Delta_6$} & \multicolumn{1}{c}{$\Delta_{7}$} & \multicolumn{1}{c}{$\Delta_{8}$} & \multicolumn{1}{c}{$\Delta_{9}$} & \multicolumn{1}{c}{$\Delta_{10}$} & \multicolumn{1}{c}{$\Delta_{2,2}$} & \multicolumn{1}{c}{$\Delta_{2,4}$} & \multicolumn{1}{c}{$\tilde{\varepsilon}_M^{(2)}$} \\
\hline
$8$ & $9$ & $-2.334800709[-8]$ & $5.173664122[-12]$ & $1.526311300[-13]$ & $-1.913660526[-13]$ & $2.781736853[-17]$ & $1.251370573[-15]$ & $-2.145865626[-13]$ & $3.345855758[-16]$ & $-2.334308513[-8]$ \\

$8$ & $10$ & $-1.838235294[-8]$ & $5.224044746[-12]$ & $1.752071911[-13]$ & $-2.471304322[-13]$ & $3.987496852[-17]$ & $1.963466727[-15]$ & $-1.932773046[-13]$ & $3.895846038[-16]$ & $-1.837739170[-8]$ \\

$8$ & $11$ & $-1.456786192[-8]$ & $4.760021559[-12]$ & $1.704614272[-13]$ & $-2.553991529[-13]$ & $4.354010146[-17]$ & $2.252894181[-15]$ & $-1.657168085[-13]$ & $3.829881106[-16]$ & $-1.456334988[-8]$ \\

$8$ & $12$ & $-1.166440321[-8]$ & $4.167836381[-12]$ & $1.554842362[-13]$ & $-2.419984846[-13]$ & $4.273504994[-17]$ & $2.283754371[-15]$ & $-1.395660353[-13]$ & $3.517402860[-16]$ & $-1.166045878[-8]$ \\

$8$ & $13$ & $-9.445800063[-9]$ & $3.590747218[-12]$ & $1.377736412[-13]$ & $-2.201643057[-13]$ & $3.984842221[-17]$ & $2.178487255[-15]$ & $-1.169776644[-13]$ & $3.130703205[-16]$ & $-9.442406153[-9]$ \\

$8$ & $14$ & $-7.735275328[-9]$ & $3.076417641[-12]$ & $1.204819601[-13]$ & $-1.962892057[-13]$ & $3.617741254[-17]$ & $2.011397858[-15]$ & $-9.816408782[-14]$ & $2.745701405[-16]$ & $-7.732370560[-9]$ \\

$8$ & $15$ & $-6.402171005[-9]$ & $2.634880261[-12]$ & $1.048057280[-13]$ & $-1.732807491[-13]$ & $3.238247545[-17]$ & $1.823817835[-15]$ & $-8.271351921[-14]$ & $2.392849229[-16]$ & $-6.399685218[-9]$ \\
\end{tabular}
\end{ruledtabular}
}
\end{table}
\end{turnpage}